\documentclass[journal,draftcls,onecolumn,12pt,twoside]{IEEEtranTCOM}

\normalsize

\ifCLASSINFOpdf

\else

\fi

\usepackage[english]{babel}
\usepackage{amsmath}		
\usepackage{graphics}		
\usepackage{psfrag}
\usepackage{setspace}		
\usepackage{longtable}          
\usepackage{amssymb,amscd,latexsym,dsfont}
\usepackage{float,color,graphicx,subfigure,balance}
\usepackage{multirow,multicol}
\usepackage{comment,cite}
\usepackage{enumerate}
\usepackage{balance}
\usepackage{stfloats}
\usepackage{tikz}
\usepackage{array,arydshln}
\usepackage{cite}
\definecolor{refkey}{rgb}{1,0.5,0} 
\definecolor{labelkey}{rgb}{1,0.5,0}
\usepackage{glossaries}
\glossarystyle{listdotted}
\makeglossaries
\renewcommand*{\CustomAcronymFields}{%
  name={\the\glsshorttok},
  description={\the\glslongtok},
  first={\noexpand\emph{\the\glslongtok}\space(\the\glsshorttok)},%
  firstplural={\noexpand\emph{\the\glslongtok\noexpand\acrpluralsuffix}\space(\the\glsshorttok)},%
  text={\the\glsshorttok},%
  plural={\the\glsshorttok\noexpand\acrpluralsuffix}%
}

\newcommand{\tr}[1]{\textrm{#1}}
\newcommand{\mb}[1]{\mathbf{#1}}

\newcommand{\mc}[1]{\mathcal{#1}}
\newcommand{\mn}[1]{{\scalebox{.45}{${\mathcal{{#1}}}$}}}
\newcommand{\mf}[1]{\mathsf{#1}}
\newcommand{\ms}[1]{\mathds{#1}}

\newcommand{\bset}[1]{\bigl\{#1\bigr\}}
\newcommand{\Bset}[1]{\Bigl\{#1\Bigr\}}

\newcommand{\cd}{\cdot}

\newcommand{\vrho}{\varrho}

\newcommand{\PR}[1]{\Pr\left\{#1\right\}}       
\newcommand{\pdf}{\tr{p}}       
\newcommand{\cdf}{\tr{F}}       

\newcommand{\secref}[1]{Section~\ref{#1}}

\newcommand{\figref}[1]{Figure~\ref{#1}}

\newcommand{\corref}[1]{Corollary~\ref{#1}}

\newcommand{\appref}[1]{Appendix~\ref{#1}}

\newcommand{\Pout}{P_{\tr{\tiny{out}}}}

\newcommand{\po}[1]{{P}_{#1}^{\mn{SD}}}
\newcommand{\pt}[1]{{P}_{#1}^{\mn{SR}}}
\newcommand{\ptr}[2]{{P}_{{#1},{#2}}^{\mn{SRD}}}
\newcommand{\tpz}[1]{\tilde{P}_{#1}}
\newcommand{\tpo}[1]{{\tilde{P}}_{#1}^{\mn{SD}}}
\newcommand{\tpt}[1]{{\tilde{P}}_{#1}^{\mn{SR}}}
\newcommand{\tptr}[2]{{{\tilde{P}}}_{{#1},{#2}}^{\mn{SRD}}}
\newcommand{\ttpz}[1]{\check{P}_{#1}}
\newcommand{\ttpo}[1]{{\check{P}}_{#1}^{\mn{SD}}}
\newcommand{\ttpt}[1]{{\check{P}}_{#1}^{\mn{SR}}}
\newcommand{\ttptr}[2]{{{\check{P}}}_{{#1},{#2}}^{\mn{SRD}}}

\newcommand{\cpo}{{\check{P}}^{\mn{SD}}}
\newcommand{\cpt}{{\check{P}}^{\mn{SR}}}
\newcommand{\cptr}{{{\check{P}}}^{\mn{SRD}}}

\newcommand{\tJ}{\tilde{J}}
\newcommand{\cJ}{\check{J}}
\newcommand{\tPout}{\tilde{{P}}_{\tr{\tiny{out}}}}

\newcommand{\xp}{X'}
\newcommand{\yp}{Y'}

\newcommand{\D}{\tr{D}}
\newcommand{\tD}{\tilde{\D}}

\renewcommand{\iota}{\nu}
\newcommand{\mnR}{{\mn{R}}}

\newcommand{\mnD}{{\mn{D}}}
\newcommand{\mcR}{{\mc{R}}}
\newcommand{\mcS}{{\mc{S}}}
\newcommand{\mcD}{{\mc{D}}}
\newcommand{\Lf}{{L^{\ast}}}
\newcommand{\Ls}{{L}}

\newcommand{\mnn}[1]{{\scalebox{.5}{${{{#1}}}$}}}
\newcommand{\plo}[2]{P_{#1}^\mnn{#2}({l})}

\newcommand{\Kp}{{\acute{K}}}

\newcommand{\Mp}{{\acute{M}}}
\newcommand{\Kpp}{{\mathbb{K}}}
\newcommand{\Mpp}{{\mathbb{M}}}

\newcommand{\alfaA}{\kappa_1}
\newcommand{\alfaB}{\kappa_2}
\newcommand{\alfaC}{\kappa_3}

\newtheorem{corollary}{Corollary}

\newtheorem{proposition}{Proposition}

\def \scalevalueS {.81}

\def \widthRatio {0.62}
\renewcommand{\baselinestretch}{1.47}
\allowdisplaybreaks
\begin{document}
\newacronym{csi}{CSI}{channel state information}
\newacronym{cqi}{CQI}{channel quality indicator}
\newacronym{ack}{ACK}{acknowledgement}
\newacronym{arq}{ARQ}{automatic repeat request}
\newacronym{awgn}{AWGN}{additive white Gaussian noise}
\newacronym{cc}{CC}{Chase combining}
\newacronym{dp}{DP}{dynamic programming}
\newacronym{fec}{FEC}{forward error correction}
\newacronym{harq}{HARQ}{hybrid automatic repeat request}
\newacronym{hspa}{HSPA}{high speed packet access}
\newacronym{iid}{i.i.d.}{independent and identically distributed}
\newacronym{ir}{IR}{incremental redundancy}
\newacronym{lte}{LTE}{long term evolution}
\newacronym{mdp}{MDP}{markov decision process}
\newacronym{mrc}{MRC}{maximal-ratio combining}
\newacronym{nack}{NACK}{negative acknowledgement}
\newacronym{pdf}{pdf}{probability density function}
\newacronym{wimax}{WiMax}{worldwide interoperability for microwave access}
\newacronym{3gpp}{3GPP}{3rd generation partnership project}
\newacronym{ofdm}{OFDM}{orthogonal frequency-division multiplexing}
\newacronym{wlan}{WLAN}{wireless local area network}
\newacronym{gsm}{GSM}{global system for mobile communications}
\newacronym{edge}{EDGE}{enhanced data \gls{gsm} environment}
\newacronym{amc}{AMC}{adaptive modulation and coding}
\newacronym{snr}{SNR}{signal to noise ratio}
\newacronym{sinr}{SINR}{signal to interference and noise ratio}
\newacronym{mi}{MI}{mutual information}
\newacronym{acmi}{ACMI}{accumulated \gls{mi}}
\newacronym{nacmi}{NACMI}{normalized \gls{acmi}}
\newacronym{cdi}{CDI}{channel distribution information}
\newacronym{latr}{LATR}{long-term average transmission rate}
\newacronym{rtr}{RTR}{round transmission rate}
\newacronym{pomdp}{POMDP}{Partially Observable Markov Decision Process}
\newacronym{fd}{FD}{full-duplex}
\newacronym{hd}{HD}{half-duplex}
\newacronym{td}{TD}{Time Division}
\newacronym{tdma}{TDMA}{time division multiple access}
\newacronym{mac}{MAC}{Media Access Control}
\newacronym{uwb}{UWB}{Ultra Wideband}
\newacronym{ieee}{IEEE}{institute of electrical and electronics engineers}
\newacronym{dB}{dB}{decibel}
\newacronym{cdf}{cdf}{cumulative density function}
\newacronym{min}{Min.}{minimum}
\newacronym{med}{Med.}{median}
\newacronym{avg}{Avg.}{average}
\newacronym{ul}{UL}{up-link}
\newacronym{dl}{DL}{down-link}
\newacronym{app}{APP}{a-posteriori probability}
\newacronym{logmap}{LogMAP}{log maximum a-posteriori}
\newacronym{llr}{LLR}{log-likelihood ratio}
\newacronym{ue}{UE}{user equipment}
\newacronym{qos}{QoS}{quality of service}
\newacronym{5g}{5G}{5\textsuperscript{th} generation mobile networks}
\newacronym{4g}{4G}{4\textsuperscript{th} generation mobile networks}
\newacronym{tti}{TTI}{transmission time interval}
\newacronym{rrm}{RRM}{radio resource management}
\newacronym{mmib}{MMIB}{mean mutual information per bit}
\newacronym{dsi}{DSI}{decoder state information}
\newacronym{tb}{TB}{transport block}
\newacronym{tbs}{TBS}{transport block size}
\newacronym{cb}{CB}{code block}
\newacronym{cbs}{CBS}{code block size}
\newacronym{prb}{PRB}{physical resource block}
\newacronym{rb}{RB}{resource block}
\newacronym{bler}{BLER}{block error rate}
\newacronym{crc}{CRC}{cyclic redundancy check}
\newacronym{tdd}{TDD}{time division duplex}
\newacronym{fdd}{FDD}{frequency division duplex}
\newacronym{embb}{eMBB}{enhanced mobile broadband}
\newacronym{mcc}{MCC}{mission critical communication}
\newacronym{mmc}{MMC}{massive machine communication}
\newacronym{mtc}{MTC}{machine type of communication}
\newacronym{mmtc}{mMTC}{massive \gls{mtc}}
\newacronym{umtc}{uMTC}{ultra-reliable \gls{mtc}}
\newacronym{urllc}{URLLC}{ultra-reliable low latency communication}
\newacronym{rtt}{RTT}{round trip time}
\newacronym{rs}{RS}{reference symbols}
\newacronym{kpi}{KPI}{key performance indicator}
\newacronym{tx}{Tx}{transmitter node}
\newacronym{rx}{Rx}{receiver node}
\newacronym{cran}{C-RAN}{centralized radio access network}
\newacronym{rru}{RRU}{remote radio unit}
\newacronym{bbu}{BBU}{baseband unit}
\newacronym{fhd}{FHD}{fronthaul delay}
\newacronym{cch}{CCH}{control channel}
\newacronym{saw}{SAW}{stop-and-wait}
\newacronym{qci}{QCI}{\gls{qos} class identifier}
\newacronym{gbr}{GBR}{guaranteed bit rate}
\newacronym{mbr}{MBR}{maximum bit rate}
\newacronym{ngbr}{non-GBR}{non-\gls{gbr}}
\newacronym{arp}{ARP}{allocation and retention priority}


\title{Opportunistic Relaying without Tx-CSI:\\ Optimizing Variable-Rate HARQ}


\author{
    \IEEEauthorblockN{Saeed R. Khosravirad, \textsl{Member, IEEE,} Leszek Szczecinski, \textsl{Senior Member, IEEE,} \\ and Fabrice Labeau, \textsl{Senior Member, IEEE}}
\thanks{S. R. Khosravirad is with Nokia Bell-Labs; He was with the Department of Electrical and Computer Engineering, McGill University, Montreal, Canada at the time this work was carried out [e-mail: saeed.khosravirad@mail.mcgill.ca]}
\thanks{F. Labeau is with the Department of Electrical and Computer Engineering, McGill University, Montreal, Canada [e-mail: fabrice.labeau@mcgill.ca].}
\thanks{L. Szczecinski is with INRS, Montreal, Canada [e-mail: leszek@emt.inrs.ca].}
\thanks{This work was presented in part at the IEEE Wireless Communications and Networking Conference (WCNC'14), Istanbul, Turkey, Apr. 2014.}
}

\maketitle
\vspace{-34pt}

\begin{abstract}

We analyze the opportunistic relaying based on \gls{harq} transmission over the block-fading channel with absence of \gls{csi} at the transmitter nodes. We assume that both the source and the relay are allowed to vary their transmission rate between  the HARQ transmission rounds. We solve the problem of throughput maximization with respect to the transmission rates using  double-recursive \gls{dp}. Simplifications are also proposed to diminish the complexity of the optimization.  The numerical results confirm that the variable-rate HARQ can increase the throughput significantly comparing to its fixed-rate counterpart. We extend the analysis to a network of $M$ relay nodes and present closed form representations of throughput for the general problem.
\end{abstract}

\section{Introduction}
\label{Sec:Introduction}

The ideas of cooperation and relaying were introduced as possible solutions to the  continuously increasing  demand on reliable data transmission that the wireless communication is facing \cite{Sendonaris:2003}. Many previous works analyzed the relay-based cooperative communications assuming the transmitters know the \gls{csi} before the transmission occurs, e.g., \cite{Host:2005,Tuninetti:2011}. In particular, it was shown in \cite{Host:2005,Laneman:2004} that, in order to increase the achievable transmission rate, the source and the relay should optimize the shares of their transmission time. In other words, the transmission \emph{rates} must vary from one transmission to another. In this work we focus on   the opportunistic relaying scenario and present a low-complexity optimization framework for finding the throughput maximizing set of transmission rates for a truncated \gls{harq} process. We will further focus on throughput-optimal rate allocation solutions of one relay collaborating with the source to deliver the message to the destination.


A more challenging  scenario is the one in which the \gls{csi} is unknown to the transmitter and then, to deal with unavoidable outage, \gls{harq} may be used to increase the reliability of the transmission. Indeed, \gls{harq} is  considered as a solution to make  relay networks practical \cite{Zhao:2005,Iannello:2009} and relay-based \gls{harq} has received a considerable interest showing significant improvements over conventional multi-hop relaying protocols \cite{Agustin:2005,Stanojev:2006,Tomasin:2007,Zheng:2010,Zimmermann:2005}. In a relay-based \gls{harq} the transmission is done in two phases: \emph{broadcasting}, where the source transmits, and the \emph{relaying} where the relay takes the transmission over if the destination does not manage to decode the message. The number of transmission rounds in both phases is random but their total number may be limited (truncated).

We follow this line of thought here but, unlike most of previous works, e.g., \cite{Zhao:2005,Byun:2008,Hu:2011}, we allow the relay and the source to optimize their transmission rates, which is similar in spirit to the work of \cite{Host:2005}. In our scheme, we optimize the rates used by the source and by the relay in each \gls{harq} transmission rounds. By analogy to \cite{Host:2005}, we call the resulting scheme a \emph{variable-rate} \gls{harq}. In a way, we bridge the results of rate-optimized known-\gls{csi} transmission of \cite{Host:2005} with those of \cite{Zhao:2005}, where the rates are not optimized and the \gls{csi} in unknown.

A similar problem has been already addressed in \cite{Byun:2009} where the rates vary on a per-transmission phase (broadcasting/relaying) basis but do not change throughout the transmission rounds in each phase. \cite{Khosravirad:2013} addressed this issue, optimizing all the rates  but assumed existence of a multi-bit feedback conveying information about the decoder state to the transmitting party. Our work has the same assumptions as \cite{Khosravirad:2013} but removes the need for multi-bit feedback and all rates vary solely as a function of the index of the transmission round for each transmission phase.

We build  partially on the results of  \cite{Leszek:2013} obtained for a point-point transmission which applied \gls{dp} optimization using simplified relationship between outage events. Here, we have to deal with the additional difficulty of having two transmission phases; this not only increases the number of rates to be optimized but also makes the relationship between outage events much  more involved.

Our contributions may be  summarized as follows:
\begin{enumerate}
\item   We introduce a variable-rate cooperative scheme based on the conventional single-bit feedback \gls{harq} transmission with opportunistic relaying. We show that the presented scheme  significantly  outperforms the   fixed-rate cooperative \gls{harq}.
\item   The non-convex optimization problem is modified so that it can be solved using doubly-recursive (or \emph{nested loop}) \gls{dp}. We further propose two simplifications to further diminish the complexity of the optimization.
\item   We compare the proposed optimization techniques to the simple alternative, which is based on random selection of the initialization point, and we show that the proposed \gls{dp} based optimization technique  provides solution very close to the best we could obtain through a much more complex alternative approach.
\item Finally, we show  numerical examples of the throughput in various topologies, which illustrates the advantages of the proposed variable-rate when comparing to the fixed-rate transmission as well as  the penalty with respect to the \gls{csi}-aware transmission.
\end{enumerate}

The rest of the paper is organized as follows, \secref{Sec:Model} explains the problem and describes the system models, \secref{Sec:THcalc} shown how to calculate the outage probability and the throughput for different network scenarios, \secref{THOptimization} and \secref{Sec:DPDesign} explain how we can cast the throughout optimization into a recursive \gls{dp} problem, \secref{Sec:NumResults} presents the numerical results while \secref{Sec:Complexity} discusses the complexity of the optimization problem.  \secref{Sec:Conclusions} concludes the work.


\section{Problem Setup for One Relay Network Scenario}\label{Sec:Model}

The cooperative communication model considered in this paper consists of three half-duplex nodes: the \emph{source} $\mc{S}$, the \emph{destination} $\mc{D}$, and  the \emph{relay} $\mc{R}$, as shown schematically in \figref{Fig:Topology}.

\subsection{Relaying Protocol}
At the begining,  the node  $\mc{S}$ is the only party in the network that has the message. The goal is to deliver the message to $\mc{D}$, possibly with the help of $\mc{R}$. We assume an error-free feedback network exists between all nodes. The feedback message is a single-bit \gls{ack} or \gls{nack} which only identifies  the success or the failure of the decoding. The transmission terminates if decoding is successful at node $\mc{D}$ in which case the node $\mc{S}$ starts transmitting the next packet from its buffer. The communication starts with the node $\mc{S}$  broadcasting the message to the other two nodes until  either node $\mc{R}$ or node $\mc{D}$  successfully decodes the message; this is the \emph{broadcasting phase}. In the case the node $\mc{R}$ decodes the message before the node $\mc{D}$ does, $\mc{R}$ starts forwarding it to $\mc{D}$ (decode-and-forward relaying) and $\mc{S}$ goes silent; this is the \emph{relaying phase}. The total number of transmission rounds by $\mc{S}$ or by  $\mc{R}$ is limited to $K$, that is, we consider  truncated \gls{harq}.

\subsection{Signal Model}

The received signal in the $k$th transmission round $(1\leq k \leq K)$ at node $\mf{b} \in \{\mc{R},\mc{D}\}$ while transmitted from node $\mf{a} \in \{\mc{S},\mc{R}\}$ is given by
\begin{align}\label{Eq:ReceivedSignal}
\mb{y}^\mf{b}_k = \sqrt{\gamma^\mf{ab}_k} \mb{x}^\mf{a}_k + \mb{z}^\mf{b}_k,
\end{align}
where  $\mb{z}_k$ is the zero mean unit variance complex Gaussian noise of the channel at  the $k$th transmission, message symbols and the noise of the channel are assumed to have unit variance, and $\gamma^\mf{ab}_k$ is the \gls{snr},  which is assumed to be perfectly known at node $\mf{b}$. The channel is  block-fading, that is $\gamma^\mf{ab}_k$ are modelled as \gls{iid} random variables varying from one transmission to another. This idealization is compatible with the assumption used in \cite{Host:2005}\cite{Zhao:2005}.

\setlength\dashlinedash{0.3pt}
\setlength\dashlinegap{2pt}
\begin{figure}[t]
\begin{center}
\psfrag{a}[c][c][\scalevalueS]{\begin{tabular}{@{}c@{}}
   Source ($\mc{S}$)\\\hdashline
   Encoder
\end{tabular}}
\psfrag{b}[c][c][\scalevalueS]{Relay ($\mc{R}$)}
\psfrag{c}[c][c][\scalevalueS]{\begin{tabular}{@{}c@{}}
   Destination ($\mc{D}$)\\\hdashline
   Decoder
\end{tabular}}
\psfrag{d}[c][c][\scalevalueS]{\begin{tabular}{@{}c@{}}
   Channel\\\hdashline
   $\pdf(y,y_1|x,x_1)$
\end{tabular}}
\psfrag{1}[b][b][\scalevalueS]{$M$}
\psfrag{2}[b][b][\scalevalueS]{$\mb{x}$}
\psfrag{3}[r][l][\scalevalueS]{$\mb{y}_1$}
\psfrag{4}[l][r][\scalevalueS]{$\mb{x}_1$}
\psfrag{5}[b][b][\scalevalueS]{$\mb{y}$}
\psfrag{6}[b][b][\scalevalueS]{$\acute{M}$}
\includegraphics[width=\widthRatio\linewidth,keepaspectratio]{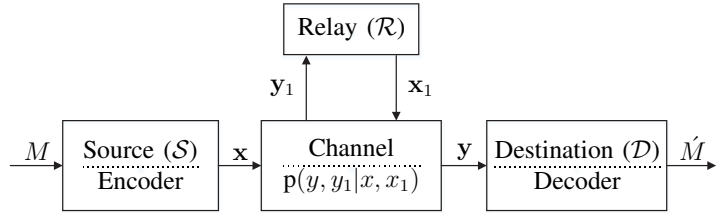}
\caption{Relay channel.}\label{Fig:Topology}
\end{center}
\end{figure}

\subsection{Variable-Rate Incremental Redundancy Transmission}

The packet of $N_\tr{b}$ information bits of message $M$ is encoded into a codeword $\mb{x}$ with  $N_\tr{s}$ symbols $x_1, x_2, \ldots, x_{N_\tr{s}}$. The symbols in the codewords  are drawn randomly from  the capacity-achieving distribution  (which is Gaussian distribution for Rayleigh block-fading channel) and the codebook is known to all the nodes.

The ARQ process starts  in broadcasting phase, when  only $\mc{S}$ has the message.  A sub-codeword $\mb{x}_1$, including only a subset of $N^{\mn{S}}_{\tr{s},1}$ symbols of the codeword, is broadcasted. Feedback messages from both $\mc{D}$ and $\mc{R}$ are sent back to $\mc{S}$ right after decoding the received packet. In case of decoding failure at both  $\mc{D}$ and $\mc{R}$, the next sub-codeword $\mb{x}_2$ of  $N^{\mn{S}}_{\tr{s},2}$ symbols is generated and broadcasted.  All the sub-codewords are assumed to be disjoint parts of the same codeword. Broadcasting phase stops whenever the message is decoded at either $\mc{D}$ or $\mc{R}$ node or if the maximum number of transmission rounds $K$ is reached.

At any time $k$ ($k<K$), if $\mc{R}$  successfully decodes the message  then we get into  relaying phase  where  $\mc{R}$ starts transmitting the message. A sub-codeword $\mb{x}_{k+1}$ of $N^{\mn{R}}_{\tr{s},k+1}$ symbols will then be generated and relayed to $\mc{D}$. This  will continue for $K-k$ rounds or will stop if $\mc{D}$ decodes the message. For simplicity, we do not impose any constraints on the size of the transmitted codewords.

For  notational convenience we use normalized sub-codeword length (redundancy) $\rho_{k} = N_{\tr{s},k}/N_\tr{b}$. It might be noticed that the redundancy $\rho_k$  has the measure of the number of channel uses per information bit and is equal to the inverse of the $k$th transmission rate $\rho_k = 1/R_k$. In \cite{Ilmu:2010} the special case of $\rho_k \equiv \rho_1, \forall k$ has been analyzed while in this paper we assume variable rate (redundancy) transmission.

Denoting the set of transmission rates for node $\mc{S}$ by $\pi^{\mn{S}}$, it is a vector of length $K$ (the maximum number of transmission rounds). Moreover, depending on the time $l$ at which the relay node decodes the message successfully and the system transits from broadcasting phase  to relaying phase, a different set of transmission rates $\pi_l^{\mn{R}}$ for the $K-l$ remaining transmissions from relay node are employed. Hence, altogether, there are $\frac{K(K+1)}{2}$ variables to be optimized. So, inputs to this problem are a set of transmission rates denoted as
\begin{align}\label{Eq:RhoSetS} \allowdisplaybreaks
\pi^{\mn{S}} &: \{\rho_k^{\mn{S}} | 1 \leq k \leq K\}  \\
\pi_l^{\mn{R}}& : \{\rho_{l,k}^{\mn{R}} | l < k \leq K\}, \quad 1 \leq l < K. \label{Eq:RhoSetR}
\end{align}

By $\pi$ we denote the rate policy of an \gls{harq}, i.e., $\pi = \{\pi^{\mn{S}},\pi_l^{\mn{R}}|1 \leq l < K\}$.

\subsection{Normalized Accumulated Mutual Information}

We denote the normalized mutual information  between two nodes $\mf{a}$ and $\mf{b}$ at time $k$  by $\iota_k^{\mf{ab}} = C(\gamma_k^{\mf{ab}}) \cdot \rho_k^{\mf{a}}$. Because we assume  Gaussian distributed symbols, the mutual information per channel use (symbol) at the decoder of node $\mf{b}$ is equal to $C(\gamma_k^{\mf{ab}}) \triangleq C^\mf{ab}_k = \log_2(1+\gamma^\mf{ab}_k)$.   From this definitions it follows that the \gls{nacmi} at the decoder of $\mc{D}$ at the end of the transmission time $k$, denoted by $I^{\mn{D}}_k$, for two example cases would be as follows:
\begin{itemize}
\item In case that relay node decodes the message at the $l$th attempt $I_k^{\mn{D}} = \sum_{m = 1}^l \iota_m^{\mn{SD}} + \sum_{m = l+1}^k \iota_{l,m}^{\mn{RD}}$,
\item In case that relay node doesn't decodes the message up to the time $k$ then, $I_k^{\mn{D}} = \sum_{m = 1}^k \iota_m^{\mn{SD}}$.
\end{itemize}

As  discussed in \cite{Caire:2001}, the probability of decoding  failure can be arbitrarily small  if the average \gls{nacmi} at the decoder of the receiver node is larger than one. Therefore, in  broadcasting phase  $I^{\mn{R}}_k < 1$ (\gls{nacmi} at the decoder of $\mc{R}$ aafter $k$ transmission attempts), and in relaying phase  $I^{\mn{R}}_k \geq  1$ (noting that at both phases  $I^{\mn{D}}_k < 1$).  Also, $\gamma^{\mn{RD}} \equiv 0$ in the broadcasting phase while $\gamma^{\mn{SD}} \equiv 0$ and $\gamma^{\mn{SR}} \equiv 0$ during the relaying phase. The transmission process stops as soon as $I^{\mn{D}}_k \geq 1$ or $k=K$.

With this notation, the outage, i.e., the event of not delivering the message to the destination, has the probability given by
\begin{align}\label{Eq:Pout}
\Pout = \Pr \{I^{\mn{D}}_K < 1 \}.
\end{align}



\section{Throughput Calculation}\label{THOptimization}
\label{Sec:THcalc}

We consider  throughput as the criterion for the optimization work in this research. We start with representing the throughput as a closed form function of the variable transmission rates (redundancies) and then will try to find the optimal transmission policies.

Based on the \textsl{reward-renewal} theorem \cite{Caire:2001}, the throughput is the ratio $\eta = \overline{N}_\tr{b}/\overline{N}_\tr{s}$ between the expected number of correctly received bits $\overline{N}_\tr{b}$ and the expected number of  channel uses $\overline{N}_\tr{s}$ used by the HARQ protocol in the $K$ transmission rounds to deliver the message packet. The number of correctly received bits for a $N_\tr{b}$-bit packet will be zero with a probability equal to  $\Pout$, or $N_\tr{b}$ with a probability of $1-\Pout$. So the throughput of the protocol is given by
\begin{align}\label{Eq:Th}
\eta = \frac{N_\tr{b}\cd (1-\Pout)}{\overline{N}_\tr{s}}.
\end{align}

\subsection{One Relay Network Scenario}

First, we present the throughput calculation problem for the  network setup that was described in \secref{Sec:Model}  where only one relay is present. We define here the probabilities of  the events which will become important when  calculating the outage probability and the throughput.
\begin{subequations}
\label{Eq:Prob.1.2.3}
\begin{equation}\label{Eq:P-one}
\po{k} \triangleq \Pr \Bset{ \sum_{i=1}^{k} \iota_i^\mn{SD} <1 }
\end{equation}
\begin{equation}\label{Eq:P-two}
\pt{k} \triangleq \Pr \Bset{\sum_{i=1}^{k} \iota_i^\mn{SR} <1 }
\end{equation}
\begin{equation}\label{P-three}
\ptr{l}{k} \triangleq \Pr \Bset{\sum_{i=1}^{l} \iota_i^\mn{SD} +\sum_{i=l+1}^{k} \iota_i^\mn{RD} <1 },
\end{equation}
\end{subequations}
where $\po{k}$ and $\pt{k}$ are  the probabilities of decoding failure, respectively at the destination and the relay, after $k$ transmissions from the source node, and $\ptr{l}{k}$ is the probability of decoding failure at the destination after  $l$ transmissions from the source node followed by $k-l$ transmissions from the relay node.

\begin{proposition}
\emph{(Throughput for One Relay Network)}
\label{Corollary}
Throughput of a variable-rate cooperative HARQ transmission protocol described in \secref{Sec:Model} can be calculated using \eqref{Eq:Th} with
\begin{align}
\Pout  =  \po{K} \cd \pt{K-1}  +  \sum_{i=1}^{K-1} \big[\pt{i-1}-\pt{i} \big]\cd \ptr{i}{K}.
\label{Eq:Pout-ALM}
\end{align}
and $\overline{N}_\tr{s}$ given in \eqref{Eq:Den-ALM}.
\end{proposition}
\begin{IEEEproof}
\appref{Ap:ThCalcAppendix}.
\end{IEEEproof}
\begin{figure*}[tb]
\normalsize
\begin{align}\label{Eq:Den-ALM}
\overline{N}_\tr{s} = \sum_{i=1}^K \rho^{\mn{S}}_i \cd \po{i-1} \cd \pt{i-1} +
 \sum_{i=1}^{K-1} \left[\pt{i-1}-\pt{i}\right].\left[\sum_{l=i+2}^K \rho^{\mn{R}}_{i,l}\cd \ptr{i}{l-1}+\rho^{\mn{R}}_{i,i+1}\cd\po{i}\right]
\end{align}
\vspace*{-20pt}
\end{figure*}


\subsection{General $M$-Relay Scenario}
\label{Sec:MrelayCase}
The throughput calculation for the variable-rate \gls{harq} transmission problem is further extended in this section to a more general scenario as shown in \figref{Fig:TopologyRelayM}, where the communication network  consists of $M$ relay nodes  $\mc{R}_m$ for $1\leq m \leq M$ other than the destination $\mcR_{M+1} = \mcD$ and the source node $\mc{R}_0 = \mc{S}$. Without loss of generality we assume that $\overline{\gamma}^{\mn{R}_m\mn{D}} < \overline{\gamma}^{\mn{R}_n\mn{D}}$ if and only if $m < n$. Assuming that every node is aware of the  distribution of its channel (i.e., physical distance) to all the other nodes in the network, the opportunistic relaying strategy  can be managed in various ways where some of them are outlined in \cite{Maham:2012}. For instance, the opportunistic relaying can be performed   by having a network of feedback channels among all the nodes. At the end of each transmission attempt, the transmitting node $\mc{R}_m$ ($0 \leq m \leq M$) and the receiver relay  nodes $\mc{R}_n, \; m < n$ will be notified about the state of the decoder of all the receiving nodes (i.e., $\mc{R}_n, \; m < n$ and $\mc{D}$). Another approach to opportunistic relaying is the centralized manner where  node $\mc{D}$ is assumed to be informed about the state of all  the decoders.

\tikzstyle{rect1}=[rectangle,text centered,text=black,text width=.08\textwidth]
\tikzstyle{rect2}=[anchor=east,text=black]
\tikzstyle{circ1} = [scale =.9, circle,minimum size=28pt,fill=gray!40]
\tikzstyle{dot1} = [scale =.5,circle,minimum size=1pt,fill=gray!50]
\tikzstyle{branch} = [scale =.75,circle,minimum size=1pt,fill=black!60]
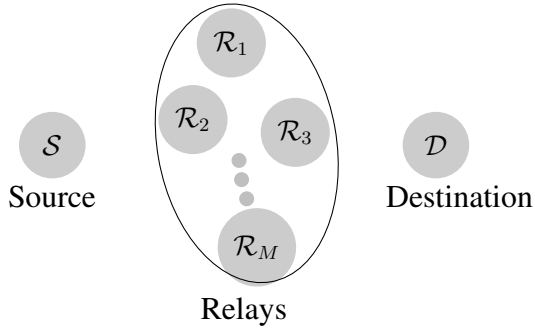
\begin{figure}
\begin{center}
\begin{tikzpicture}[scale=1.7]
  \node (n11)[circ1] at (0,0) {$\mcS$};
  \node (n12)[circ1] at (1.4,.8)  {$\mcR_1$};
  \node (n13)[circ1] at (1.1,0.2)  {$\mcR_2$};
  \node (n14)[circ1] at (1.9,.1) {$\mcR_3$};
  \node (n15)[circ1] at (1.6,-.8) {$\mcR_M$};
  \node (n16)[circ1] at (3,0)  {$\mcD$};
  \node (n17)[dot1] at (1.46,-0.12)  {};
  \node (n18)[dot1] at (1.52,-0.42)  {};
  \node (n19)[dot1] at (1.48,-0.27)  {};
  \draw [rotate=10](1.5,-.25) ellipse (.7cm and 1.1cm);
  \node (t0)[rect1] at (0,-.4) {Source};
  \node (t1)[rect1] at (1.5,-1.3) {Relays};
  \node (t2)[rect1] at (3,-.4) {Destination};
\end{tikzpicture}
\caption{Topology of the $M$-relay network.}\label{Fig:TopologyRelayM}
\end{center}
\end{figure}

Starting from $\mcR_0 = \mcS$ at time $k=0$ several paths can be taken to finally get to $\mcR_{M+1} = \mcD$ or reach the time constraint of $K$ transmission attempts. Assuming that there exists a feedback network between all nodes,  at each time $k$ only one node is active which is the best node with respect to its channel condition with node $\mcD$. A path is the set of nodes like $\{\mcS = \mcR_{l_1}, \mcR_{l_2}, \ldots, \mcR_{l_K}, \mcR_{l_\Kp} \}$, with $\Kp = K+1$, where $\mcR_{l_k} \quad 1 \leq k \leq K$ denotes the node that is active in the $k$th transmission attempt\footnote{By definition,  after a successful \gls{harq} transmission the path arrives at node $\mcD$ at the transmission attempts  $\tau$. As a result, for a success path $\mcR_{l_k} = \mcD$ for $k \geq \tau$. The arrival of the path at node $\mcD$ is equivalent to terminating the \gls{harq} process for the packet and channel will be free for starting another \gls{harq} process for the next packet.}.

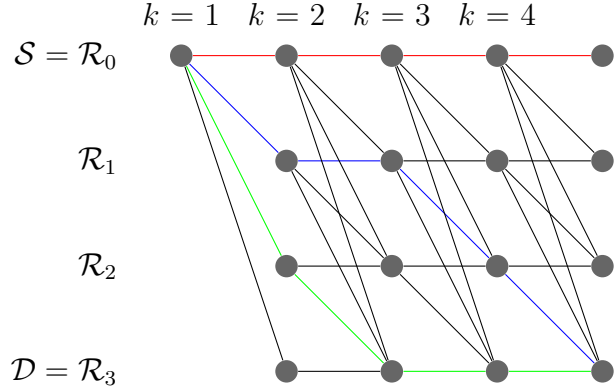
\begin{figure}
\begin{center}
\begin{tikzpicture}[scale=1.4]
  \node (n11)[branch] at (0,5) {};
  \node (n12)[branch] at (1,5)  {};
  \node (n13)[branch] at (2,5)  {};
  \node (n14)[branch] at (3,5) {};
  \node (n15)[branch] at (4,5)  {};
  \node (n22)[branch] at (1,4)  {};
  \node (n23)[branch] at (2,4)  {};
  \node (n24)[branch] at (3,4) {};
  \node (n25)[branch] at (4,4)  {};
  \node (n32)[branch] at (1,3)  {};
  \node (n33)[branch] at (2,3)  {};
  \node (n34)[branch] at (3,3) {};
  \node (n35)[branch] at (4,3)  {};
  \node (n42)[branch] at (1,2)  {};
  \node (n43)[branch] at (2,2)  {};
  \node (n44)[branch] at (3,2) {};
  \node (n45)[branch] at (4,2)  {};

    \foreach \from/\to in {n11/n32,n32/n43,n43/n44,n44/n45}
    \draw [green] (\from) -- (\to);
    \foreach \from/\to in {n11/n22,n22/n23,n23/n34,n34/n45}
    \draw [blue] (\from) -- (\to);
    \foreach \from/\to in {n11/n12,n12/n13,n13/n14,n14/n15}
    \draw [red] (\from) -- (\to);
    \foreach \from/\to in {n11/n42,n12/n23,n12/n33,n12/n43,n13/n24,n13/n34,n13/n44,n14/n25,n14/n35,n14/n45,n22/n33,n22/n43,n23/n44,n23/n24,n24/n35,n24/n45,n24/n25,n32/n33,n33/n34,n33/n44,n34/n35,n42/n43}
    \draw (\from) -- (\to);

    [scale=2,auto=left,every node/.style={circle,fill=white}]
    \node (t0)[rect1] at (-1,5.4) {};
    \node (t1)[rect1] at (0,5.4) {$k = 1$};
    \node (t2)[rect1] at (1,5.4) {$k = 2$};
    \node (t3)[rect1] at (2,5.4) {$k = 3$};
    \node (t4)[rect1] at (3,5.4) {$k = 4$};
    \node (R0)[rect2] at (-0.5,5) {$\mcS = \mcR_0$};
    \node (R1)[rect2] at (-0.5,4) {$\mcR_1$};
    \node (R2)[rect2] at (-0.5,3) {$\mcR_2$};
    \node (R3)[rect2] at (-0.5,2) {$\mcD = \mcR_3$};
\end{tikzpicture}
\caption{For a network with two relays and $K = 4$ all the paths are shown. The path shown in red, represents a failure path with the index of $(0)_{10} = (0000)_4$; respectively, the blue and green paths are success path wit indexes of, $(91)_{10} = (1123)_4$ and $(127)_10 = (1333)_4$.}\label{Fig:Path}
\end{center}
\end{figure}

Since we always have $\mcR_{l_1} = \mcS$, we denote the path by the set of node indexes after first transmission using a number of  $K$ digits in $\Mp = M+1$ base, like $(l_2  l_3  \ldots  l_\Kp)_{\Mp}$, where each digit $l_k$ denotes the active node at  $k$th  transmission attempt. Alternatively, we can denote a path by its representation in base 10, as $(l)_{10} = (l_2  \ldots  l_K)_{\Mp}$. We choose   $L$ to denote  the number of success paths (paths that end up in node $\mcD$) and $\Lf$, to denote the number of failure paths (paths that end up in any node other than $\mcD$).
Parameter $\Lf$ which shows the number of failure paths to each state can be computed as follows.
\begin{align}
\Lf = \sum_{m=0}^{M-1}L_{m,K}
\end{align}
where $L_{m,K}$ is the number of  paths that end up at node $\mcR_m$ after  $K$ transmissions and can be computed using the following recursive rule of $L_{m,k} = \sum_{i=0}^{m}L_{i,k-1}$ for $1 \leq k\leq K$ and $0 \leq m \leq M$, where by definition $L_{0,k} = 1$, $\forall k$ and $L_{m,0} = 0$, $\forall m>0$. A success path is a path that ends at node $\mcR_{M+1}$ (or $\mcD$) after $k \in {1, 2, ..., K}$ transmissions. Number of success paths $\Ls$, can be computed as $\Ls = \sum_{k=1}^{K}\Ls_{M,k}$, where $\Ls_{M,k}$ is the number of path that end up in $\mcD$ at time $k$.
Following this definition, we can assume $\Lf + \Ls$ number of events, each of them corresponding to going in one of the possible paths (either failure path or success path) in the \gls{harq} process. The following corollary states that these events are disjoint.
\begin{corollary}
\emph{(Disjoint Events)}
\label{Corollary}
Events related to each path, either a failure path or a success path, are disjoint.
\textnormal{}
\end{corollary}

\begin{proof}
Let's assume two different  paths $(l_1, ..., l_{k})$ and $(\acute{l}_1, ..., \acute{l}_{k})$ (i.e., two path that diverged from each other in one of the nodes at a time $\tau$ although they may converge back to the same point again). 
Then there  exists a $0 \leq \tau < k$ where $l_j = \acute{l}_j$ for all $1 \leq j \leq \tau$ and $l_j \neq \acute{l}_j$ for $ j = \tau + 1$. From the definition of  the opportunistic relaying presented above, the probability of being in two different nodes at any time is equal to zero. Therefore, the event of the two  paths are disjoint.
\end{proof}

The transmission redundancy from node $\mcR_m$ at time $k$ while going through path $l$, can be denoted by $\rho^m_{l,k} $. We denote the set of $\rho$ values for a path $l$  by $\vrho(l) = \{ \rho^{l_1}_{l,k}, \ldots , \rho^{l_k}_{l,k} \} $. For the proposed relaying protocol a policy is the union of all $\vrho(l)$ sets as $\pi = \bigcup_{l}\vrho(l)$. The number of different $\rho$ values available for node $\mcR_m$ to chose from for the transmission attempt $k$, where $1 <k \leq K$, is at most equal to the number of different path that end up at the node after $k-1$ transmissions, which is equal to  $L_{m,k-1} = \sum_{i = 0}^{m} L_{i,k-2}$, and as  a result the dimension of $\pi$ is bound as follows.
\begin{align}
\dim(\pi) \leq \sum_{k = 2}^{K} \sum_{m = 0}^{M-1} L_{m,k-1} + L_{0,0}  = \sum_{k = 2}^{K} \sum_{m = 0}^{M-1} \sum_{i = 0}^{m} L_{i,k-2} +1
\end{align}

For instance, $K = 3$ transmissions with a network of $M = 2$ relay nodes, gives a dimension of $20$ to the policy set $\pi$. This number grows rapidly with $K$ and $M$ as shown in the table below for a few example cases.
\begin{table}[h]
\vspace{-10pt}
\begin{center}
\begin{tabular}{cc|c|c|c|c|c}
\cline{3-6}
& & \multicolumn{4}{ c| }{$\Kp$} \\ \cline{3-6}
& & 2 & 3 & 4 & 8 \\ \cline{3-6} \cline{1-6}
\multicolumn{1}{ |c  }{\multirow{4}{*}{$M$} } &
\multicolumn{1}{ |c|| }{1} & 3 & 6 & 10 & 36 &     \\ \cline{2-6}
\multicolumn{1}{ |c  }{}                        &
\multicolumn{1}{ |c|| }{2} & 4 & 10 & 20 & 120 &     \\ \cline{2-6}
\multicolumn{1}{ |c  }{}                        &
\multicolumn{1}{ |c|| }{4} & 6 & 21 & 56 & 792 &     \\ \cline{2-6}
\multicolumn{1}{ |c  }{}                        &
\multicolumn{1}{ |c|| }{10}& 12 & 78 & 364 & 31824 &     \\ \cline{1-6}
\end{tabular}
\end{center}
\caption{Dimension of policy set $\pi$.}
\label{table:name}
\vspace{-40pt}
\end{table}

The number of paths, which is equal to the number of $\Kp$ digit numbers in base $\Mp$ where the highest order digit is equal to $0$ and  a higher order digit is always less or equal to the lower order digit, can be found as follows.
\begin{align}
\Ls + \Lf =  \sum_{k = \max(0,\Kpp-\Mpp)}^{\Kpp-1} \dbinom{\Mpp}{\Kpp-k}\dbinom{\Kpp-1}{\Kpp-k-1}   = \sum_{k = 0}^{\Kpp-1} \dbinom{\Mpp}{\Kpp-k}\dbinom{\Kpp-1}{\Kpp-k-1}
\end{align}
where $\Kpp = K$ and $\Mpp = M+2$.

We define the probability of decoding failure at node $\mc{\alpha}$ while transmission goes through  a path $l$ as follows.
\begin{align}\label{Eq:PLOalpha}
\plo{k}{\alpha}\triangleq \PR{ \sum_{i=1}^{k}I_i^{\mnR_\mnn{l_{i}}{\alpha}} < 1 }
\end{align}

\begin{proposition}
\emph{(Throughput for $M$-Relay Network)}
\label{Corollary1}
Throughput of a variable-rate cooperative \gls{harq} transmission with opportunistic relaying among $M$ relay nodes can be calculated as 
\begin{align}
\eta = \frac{1-\Pout}{\D} = \frac{1 - \sum_{l \in failure} \Pr\{E(l)\}}{\sum_{l} \Pr\{E(l)\} \cd \overline{\vrho}^l}
\label{Eq:Eta-ALMM}
\end{align}
where,
\begin{align}
\label{Eq:vrhol}
\overline{\vrho}^l = \sum_{\rho_{l,m}^k \in \vrho(l)} \rho_{l,m}^k.
\end{align}
\end{proposition}
\begin{IEEEproof}
\appref{Ap:MThCalcAppendix}.
\end{IEEEproof}


\section{Throughput Optimization}\label{Sec:DPDesign}

We discuss the throughput optimization problem for the one relay scenario although this could be extended to the $M$-relay scenario as we will discuss later in this section. The throughput   of \gls{harq} for $K$  retransmissions as introduced in \eqref{Eq:Th} has $K(K+1)/2$  optimization variables which makes it a complex optimization problem. To reduce the complexity to a reasonable order, we present a dual optimization problem  inspired by \cite{Leszek:2013} and \cite{Khosravirad:2014} and solve the optimization problem in a recursive manner which greatly  reduces the complexity of the problem.

\subsection{Dual Optimization  Problem}

We denote by $\pi = \{\pi^{\mn{S}}, \pi_l^\mn{R}\}$ the set of redundancies for a truncated cooperative \gls{harq} transmission for $1 < l < K$. Furthermore, we denote the  $\overline{N}_\tr{s}$  in \eqref{Eq:Den-ALM} by $\D(\pi)$ since it is naturally a function of the policy $\pi$. The throughput for the set of redundancies $\pi$ is
\begin{align}\label{TH.1}
\eta(\pi)  =\frac{1-\Pout(\pi)}{\D(\pi)}.
\end{align}
Denoting the maximum throughput by $\hat{\eta}$, the throughput maximization problem can be represented as
\begin{align}\label{TH.max}
\hat{\eta} = \max_\pi \eta(\pi).
\end{align}

The optimization problem above has $K(K+1)/2$  optimization variables, meaning that it has a polynomially  increasing complexity of order $K^2$. As we will see in \secref{Sec:Complexity}, the problem is not convex and the conventional gradient-based optimization is not appropriate in this case. We thus cast the problem into a recursive form using approximations. While the solution are suboptimal, the global solution of the new problem can be obtained with predefined complexity. The first step in order to have a recursive form of \eqref{TH.max} is to eliminate the  fraction. As proposed in \cite{Leszek:2013} we change the optimization  \eqref{TH.max} to the dual problem
\begin{align}
J^\lambda = \min_\pi \D(\pi)+\lambda \cdot \Pout(\pi),
\label{Eq:J.lambda}
\end{align}
where $\lambda$ is the Lagrange multiplier.

We call a redundancy set $\pi$  \textsl{degenerate} if it guarantees zero redundancy transmission, and consequently  $\Pout(\pi)=1$ (this is the same as saying that $\pi$ is degenerate if and only if  $\D(\pi) = 0$ which happens if and only if  $\rho^{\mn{S}}_k = 0 \quad \forall k$). We also call $\pi$ \textsl{non-degenerate} if it is not degenerate. As proved in \cite{Khosravirad:2013}, the maximization problem \eqref{TH.max} is equivalent to finding $\lambda_\tr{th}$ for \eqref{Eq:J.lambda} which is the smallest value of $\lambda$ where a non-degenerate solution for $J^\lambda$ can be found.

\subsection{Approximate Optimization}

In order for \eqref{Eq:J.lambda} to be fashioned in a \gls{dp} recursive representation, we need to choose a term as the  ``\emph{state}'' of the recursive optimization, denoted by $S_k$, which has  the  following two conditions \cite{Visotsky:2003,Bertsekas:1995}. First, knowing the $k$th optimization parameter (redundancy variables $\rho$ in our problem) and ${S}_{k}$, the new state ${S}_{k+1}$ should be obtained. This makes it possible to optimize each of the variables separately. Second,  the probability of failure events at the end of $k$th transmission must be computed knowing ${S}_{k}$.

The probability of failure events in \eqref{Eq:Prob.1.2.3} at time $k$ depend on all the $\rho$ variables up to the time. Therefore, the problem in \eqref{Eq:J.lambda} does not have the second condition mentioned above to be cast into \gls{dp} recursive format. As already suggested in \cite{Caire:2001,Wu:2010,Leszek:2013} we choose to do some modification to the problem to overcome this issue. We  approximate the probability of  failure events  using a Gaussian approximation \cite{Sesia:2004} with two dimensional state of $S_k = (X_k,Y_k)$. For instance for $\po{k}$ in \eqref{Eq:P-one} we use  $\tpo{k}$ where
\begin{equation}\label{Eq:Pt-one-1}
\tpo{k} = \left\{\begin{matrix}
\cdf_{C^\mn{SD}}\big(\frac{1}{\rho^{\mn{S}}_k} \big), & k = 1 \vspace{3pt}\\
Q\Big(\frac{\overline{C}^\mn{SD}\cd X_k - 1}{\sigma_{C^\mn{SD}} \cd \sqrt{Y_k}} \Big), & \tr{otherwise}.
\end{matrix}\right.
\end{equation}
In \eqref{Eq:Pt-one-1}, $\overline{C}^\mf{ab} = \ms{E}_{C^\mf{ab}}\bset{C^\mf{ab}}$ and $\sigma_{C^\mf{ab}}^2 = \ms{E}_{C^\mf{ab}}\bset{{C^\mf{ab}}^2}- {\overline{C}^\mf{ab}}^2$. Also, $X_k = \sum_{l=1}^{k}\rho^{\mn{S}}_l$, $Y_k = \sum_{l=1}^{k}{\rho^{\mn{S}}_l}^2$, $\cdf_{C}(.)$ is the cumulative distribution function \gls{cdf} and $Q(x)$ is the Q-function defined as
\begin{align}
Q(x) = \frac{1}{\sqrt{2\pi}}\int_x^\infty \exp\big(\frac{-\tau^2}{2} \big)\tr{d}\tau.
\end{align}
We can define $\tpt{k}$ in the same way for the channel, putting $\mf{ab} = \mc{SR}$. Moreover, we approximate $\ptr{l}{k}$ with  $\tptr{l}{k}$ as follows.
\begin{equation}\label{Eq:Pt-one}
\tptr{l}{k} = \left\{\begin{matrix}
\cdf_{C^\mn{SD}}\big(\frac{1}{\rho^{\mn{S}}_k} \big), & k = 1 \vspace{4pt}\\
Q\Big(\frac{\overline{C}^\mn{SD}\cd X_k + \overline{C}^\mn{RD}\cd \xp_k- 1}{\sigma_{C^\mn{SD}} \cd \sqrt{Y_k} + \sigma_{C^\mn{RD}} \cd \sqrt{\yp_k}}  \Big), & \tr{otherwise}
\end{matrix}\right. ,
\end{equation}
where $\xp_k = \sum_{i = l+1}^{k}\rho^{\mn{R}}_{l,i}$ and $\yp_k = \sum_{i=l+1}^{k}{\rho^{\mn{R}}_{l,i}}^2$.

Using the approximate failure probabilities, the minimization problem in \eqref{Eq:J.lambda} becomes
\begin{align}\nonumber
\tJ^\lambda & = \min_\pi \bset{ \tD(\pi)+\lambda \cdot \tPout(\pi)} \\
& = \min_\pi \Bset{ \sum_{i=1}^{K-1} \big[\rho^\mc{S}_i \cd \tpo{i-1} \cd \tpt{i-1}\big] +  \tilde{f}_i \cd \tilde{g}^\lambda_i   + \lambda \cd \tpo{K} \cd \tpt{K-1} + \rho^\mc{S}_K \cd \tpo{K-1} \cd \tpt{K-1}},
\label{Eq:tJ.lambda}
\end{align}
where
\begin{align}
\label{Eq:fi}
\tilde{f}_i = \tpt{i-1}-\tpt{i}
\end{align}
and
\begin{align}\label{Eq:gi}
\tilde{g}^\lambda_i = \lambda\cd \tptr{i}{K} + \sum_{l=i+2}^K \rho^\mc{R}_{i,l}\cd \tptr{i}{l-1}+\rho^\mc{R}_{i,i+1}\cd\tpo{i}.
\end{align}

Clearly, a solution $\tilde{\pi}$ to \eqref{Eq:tJ.lambda} for any $\lambda$ value, is a suboptimal solution to \eqref{TH.max} (i.e., $\eta(\tilde{\pi}) \leq \hat{\eta}$), however it has the advantage of being easily found through a recursive optimization approach, even for large $K$.

\subsection{DP Recursive Optimization}
\label{Ap:DP}

The problem in \eqref{Eq:tJ.lambda} can be solved in $K$ recursive steps, where we use  two-dimensional state $S_k = (X_k,Y_k)$ to find  $\tJ_{1}^{\lambda}(X_0,Y_0)$  as presented in \eqref{Eq:tJ.ALM}. $\tJ_{1}^{\lambda}$, and $\tJ_{k}^{\lambda}(X_{k-1},Y_{k-1})$ for $1 < k < K$ and f$\tJ_{K}^{\lambda}$, are shown respectively in \eqref{Eq:tJ1}, \eqref{Eq:Jk-ALM} and \eqref{Eq:JK-ALM}.
\begin{figure*}[t]
{\fontsize{9.285pt}{10pt}\selectfont
\begin{subequations}
\label{Eq:tJ.ALM}
\begin{align}\label{Eq:tJ1}
\tJ_{1}^{\lambda}(X_0,Y_0) & = \min_{ \substack{\rho^{\mc{S}}_{1}}}  \bset{   \tJ_{2}^{\lambda} \left(X_{0}+\rho^\mc{S}_{1},Y_{0}+(\rho^\mc{S}_{1})^2 \right) + \rho^\mc{S}_{1} + f_{1}\cd V^{\lambda,1}\left(X_{0}+\rho^\mc{S}_{1},Y_{0}+(\rho^\mc{S}_{1})^2 \right)}\\
\tJ_{k}^{\lambda}(X_{k-1},Y_{k-1}) & = \min_{ \substack{\rho^{\mc{S}}_{k}}}  \bset{  \tJ_{k+1}^{\lambda} \left(X_{k-1}+\rho^\mc{S}_{k},Y_{k-1}+(\rho^\mc{S}_{k})^2 \right) + \rho^\mc{S}_{k}\cd\tpo{k-1}\cd \tpt{k-1} + f_{k}\cd V^{\lambda,k}\left(X_{k-1}+\rho^\mc{S}_{k},Y_{k-1}+(\rho^\mc{S}_{k})^2 \right)}  \label{Eq:Jk-ALM} \\
\tJ_{K}^{\lambda}(X_{K-1},Y_{K-1}) & = \min_{ \substack{\rho^{\mc{S}}_{K}}}  \bset{  \rho^\mc{S}_{K}\cd\tpo{K-1}\cd \tpt{K-1} + \lambda\cd \tpo{K}\cd\tpt{K-1} } \label{Eq:JK-ALM}
\end{align}
\end{subequations}}
\vspace*{-20pt}
\end{figure*}

The recursive optimization starts with \eqref{Eq:JK-ALM} to find the function $\tJ_{K}^{\lambda}$ and continues going backward on $k$ up to $k = 1$. The optimal value $ \tJ^{\lambda}$ can be found according to   $ \tJ^{\lambda} =  \tJ_{1}^{\lambda}(X_0,Y_0)|_{_{({X}_0,{Y}_0) = (0,0)}}$. The optimal policy $\pi^\mc{S}$ can then  be found  starting with $\rho^\mc{S}_1$   as follows with putting $(\hat{X}_0,\hat{Y}_0) = (0,0)$.
\begin{enumerate}
\item   $\tilde{\rho}^\mc{S}_1 = \arg_\rho \tJ_{1}^{\lambda}(\hat{X}_0,\hat{Y}_0)$
\item   for $k = 2 , \ldots,  K$
\begin{itemize}
\item   $\hat{X}_{k-1} = \hat{X}_{k-2} + \tilde{\rho}^\mc{S}_{k-1}$ and $\hat{Y}_{k-1} = \hat{Y}_{k-2} + (\tilde{\rho}^\mc{S}_{k-1})^2$
\item   $\tilde{\rho}^\mc{S}_k = \arg_\rho \tJ_{k}^{\lambda}(\hat{X}_{k-1},\hat{Y}_{k-1})$
\end{itemize}
\end{enumerate}

All the steps for the recursive optimization in \eqref{Eq:tJ.ALM}, are assuming a given $\tilde{V}^{\lambda,i}$ for $1 \leq i \leq K-1$, where
\begin{align}\label{Eq:Vlambda}
\tilde{V}^{\lambda,i}(\alpha,\beta) = \min_{ \substack{\rho^{\mc{R}}_{i,l}\in \pi^\mc{R}_i \\ \sum_{k=1}^i \rho^{\mc{S}}_{k} = \alpha, \; \sum_{k=1}^i (\rho^{\mc{S}}_{k})^2 = \beta}}  \bset{\tilde{g}^\lambda_i}.
\end{align}
As a result, before solving \eqref{Eq:tJ1}, we first need to complete a  pre-optimization step to compute $\tilde{V}^{\lambda,i}$. The function $\tilde{g}^\lambda_i$ can  be  optimized with respect to $\pi^\mc{R}_i$ only, if the two summations of $\sum_{k=1}^i \rho^{\mc{S}}_{k}$ and  $\sum_{k=1}^i (\rho^{\mc{S}}_{k})^2$ were given. This means that, optimization of the term $\tilde{g}^\lambda_i$  is \emph{nested} inside of the optimization function in \eqref{Eq:tJ.ALM}. 

As we show in the following, $\tilde{V}^{\lambda,i}(\alpha,\beta)$ can  be solved recursively and the results will be stored to be used  in the nested-loop minimization problem of \eqref{Eq:tJ.ALM}. Using \eqref{Eq:Vlambda} we can rewrite \eqref{Eq:tJ.lambda} as in \eqref{Eq:tJ.lambda.2}.
\begin{figure*}[t]
{\fontsize{9.6pt}{10pt}\selectfont
\begin{align}
\tJ^\lambda  =  \min_\pi \Bset{  \sum_{i=1}^{K-1} [\rho^\mc{S}_i \cd \tpo{i-1} \cd \tpt{i-1}] +  f_i \cd \tilde{V}^{\lambda,i}\Big(\sum_{k=1}^i \rho^{\mc{S}}_{k},\sum_{k=1}^i (\rho^{\mc{S}}_{k})^2 \Big) +  \lambda \cd \tpo{K} \cd \tpt{K-1} + \rho^\mc{S}_K \cd \tpo{K-1} \cd \tpt{K-1} }.
\label{Eq:tJ.lambda.2}
\end{align}}
\vspace*{-20pt}
\end{figure*}

For the minimization in \eqref{Eq:Vlambda}, we use a nested state of $\mf{s}_i = (\xp_i,\yp_i)$. This can be shown as follows:
\begin{align}\nonumber
\tilde{V}^{\lambda,i}(\alpha,\beta) = V_{i+1}^{\lambda,i}(\xp_{i+1},\yp_{i+1},\alpha,\beta)|_{_{(\xp_{i+1},\yp_{i+1}) = (0,0)}},
\end{align}
where $V_{i+k}^{\lambda,i}(\xp_{K-1},\yp_{K-1},\alpha,\beta)$ for $k = 1$, $1 < k < K-i$ and  $k = K-i$ are shown respectively in \eqref{Eq:Vi1-ALM}, \eqref{Eq:Vk-ALM} and \eqref{Eq:VK-ALM}.
\begin{figure*}[t]
{\fontsize{9.6pt}{10pt}\selectfont
\begin{subequations}
\begin{align}
V_{i+1}^{\lambda,i}(\xp_{i+1},\yp_{i+1},\alpha,\beta) & = \min_{ \substack{\rho^{\mn{R}}_{i,i+1} }}  \Bset{  \rho^\mc{R}_{i,i+1}\cd \tpo{i} + V_{i+2}^{\lambda,i}\left(\xp_{i+1}+\rho^\mc{R}_{i,i+1},\yp_{i+1}+(\rho^\mc{R}_{i,i+1})^2,\alpha,\beta \right)}\label{Eq:Vi1-ALM}
\\
V_{i+k}^{\lambda,i}(\xp_{i+k},\yp_{i+k},\alpha,\beta) & = \min_{ \substack{\rho^{\mn{R}}_{i,i+k} }}  \Bset{  \rho^\mc{R}_{i,i+k}.\tptr{i}{i+k-1} + V_{i+k+1}^{\lambda,i}\left(\xp_{i+k}+\rho^\mc{R}_{i,i+k},\yp_{i+k}+(\rho^\mc{R}_{i,i+k})^2,\alpha,\beta \right)} \label{Eq:Vk-ALM}
\\
V_{K}^{\lambda,i}(\xp_{K},\yp_{K},\alpha,\beta) & = \min_{ \substack{\rho^{\mn{R}}_{i,K}}}  \Bset{ \rho^\mc{R}_{i,K}.\tptr{i}{K-1} + \lambda\cd\tptr{i}{K}}\label{Eq:VK-ALM}
\end{align}
\end{subequations}}
\vspace*{-20pt}
\end{figure*}

This will be solved starting from \eqref{Eq:VK-ALM} and ending with \eqref{Eq:Vi1-ALM} considering $ \sum_{k=1}^i \rho^{\mn{S}}_{k} = \alpha$ and $ \sum_{k=1}^i (\rho^{\mn{S}}_{k})^2 = \beta^2 $. Then the set of $\rho^{\mn{R}}_{i,l}\quad i<l\leq K$ will be found  starting with $\rho^{\mn{R}}_{i,i+1}$ using \eqref{Eq:Vi1-ALM} with $(\xp_{i+1},\yp_{i+1}) = (0,0)$ and going up to $\rho^{\mn{R}}_{i,K}$ in \eqref{Eq:VK-ALM} recursively. The optimal throughput will then be $\eta(\tilde{\pi})$.

For the case of $M$-relay network, the recursive optimization approach introduced above can be adopted and further generalized.  Although we will not go into details of such an approach, the optimization  can be summarized as follows. Using the term \emph{cost} to assess the $\tJ$ and $\tilde{V}$ values introduced above, the cost of being at node $R_m$ at time $k$ is consisted of the cost of being at any $R_n$ where $m \leq n$ at time $k+1$. Although this cost can be minimized separately for all the possible paths like  $(l_2  l_3  \ldots  l_\Kp)_{\Mp}$ that has $l_k = m$ the same way that we optimize $\tilde{V}$ in the above. The result of these optimization can then be stored to be used later by another optimization for the time moment $k-1$. This way a recursive optimization will be formed similar to what was explained above for the one relay scenario and the optimal solution can be found in a similar recursive way.

\subsection{Simplified one dimensional state}
\label{Sec:ALM-oneD}

A simplified version of the proposed optimization can be obtained by modifying  the problem in a way that the \gls{dp} optimization  state is only one dimensional or ${S}_{k} = X_k$. The state elements  in \eqref{Eq:tJ.lambda} have  be discretized into $Q$ number of points and for a two dimensional space, which would create an $Q^2$ number of minimizations at each step.

Therefore, reducing  the dimension of the state space to one, will immediately decrease the complexity of the optimization process by reducing the number of minimizations in each step from $Q^2$ to $Q$.

We discuss the one dimensional state in this section using  Gaussian approximation by approximating the state elements as: $\sqrt{Y_k} \approx X_k$ and $\sqrt{\yp_k} \approx \xp_k$.


The failure probabilities $\po{k}$ (and similarly $\pt{k}$) and $\ptr{l}{k}$ when approximated as  functions of $X_k$ and $\xp_k$, are presented  as follows.
\begin{align}\label{Eq:tpo-oneDim}
\po{k} \approx  \ttpo{k}(X_k) = \left\{\begin{matrix}
\cdf_{C^\mn{SD}}\big(\frac{1}{\rho^\mn{S}_k} \big),  & k = 1 \vspace{4pt}\\
Q\Big(\frac{\overline{C}^\mn{SD}\cd X_k - 1}{\sigma_{C^\mn{SD}} \cd X_k} \Big) & \tr{otherwise}
\end{matrix}\right. .
\end{align}
\begin{align}\label{Eq:tptr-oneDim}
\ptr{l}{k} \approx \ttptr{l}{k}(X_l,\xp_k) = \left\{\begin{matrix}
\cdf_{C^\mn{SD}}\big(\frac{1}{\rho^\mn{S}_k} \big),  & k = 1 \vspace{4pt}\\
Q\Big(\frac{\overline{C}^\mn{SD}\cd X_l +\overline{C}^\mn{RD}\cd \xp_k - 1}{\sigma_{C^\mn{SD}} \cd X_l + \sigma_{C^\mn{RD}} \cd \xp_k} \Big), & \tr{otherwise}
\end{matrix}\right. .
\end{align}

As a result, to maximize the throughput using one-dimensional Gaussian approximation probabilities, we solve $\cJ^\lambda$ instead of $J^\lambda$, with substituting the outage probabilities in \eqref{Eq:J.lambda} with the approximated version. Then, the goal is to find the following.
\begin{align}\label{Eq:J.lambda.OneDim}
\cJ^\lambda = \cJ^{\lambda}_K(\check{X}_K),
\end{align}
where $\check{X}_K = \arg_X \min J^{\lambda}_K(X)$ and $\cJ^{\lambda}_K$ is presented in  \appref{Ap:OneDimensionalState} along with how to solve \eqref{Eq:J.lambda.OneDim}. After $\check{X}_K$ is found, the solution set $\check{\pi} = \pi(\check{X}_K)$ is created and $\eta(\check{\pi})$ can be computed using the exact throughput calculation.

\subsection{Performance Bounds}

For infinite number of  allowed  transmission rounds, the maximum achievable throughput reaches the ergodic capacity of the fading channel in a single-hop channel \cite{Caire:2001,Khosravirad:2013,Leszek:2013}. For the relay channel we also expect the maximum achievable throughput  to grow with $K$. In \cite{Khosravirad:2013}, for the same relay channel, it is shown that with $K \rightarrow \infty$ the maximum achievable throughput is bounded by  $\eta_{\mbox{\tiny{max}}}$ which can be found using Bellman's equation \cite[Chap.~3]{Bertsekas:1995}.

Moreover, the obvious lower bound of one transmission  happens when $K=1$  (also known as direct transmission lower bound for Decode-and-Forward channel). averaged  on the channel state. We denote this lower bound by $\hat{\eta}_{\mbox{\tiny{0}}}$ which can be calculated as
\begin{align}\label{Eq:etaMin}
\hat{\eta}_{\mbox{\tiny{0}}} = \max_{\rho^{\mn{S}}} \Bset{\frac{1-\Pout}{\rho^{\mn{S}}}},
\end{align}
where $\Pout = \Pr\Bset{C^{\mn{SD}}\cd \rho^{\mn{S}}<1 } = \cdf_{C^{\mn{SD}}}(\frac{1}{\rho^{\mn{S}}})$.

Capacity of the  relay channel with input  $\mb{x}$, relay input $\mb{x}_1$, output $\mb{y}$ and relay output $\mb{y}_1$ (\figref{Fig:Topology}) for an arbitrary channel given by $\pdf(y,y_1|x,x_1)$ and a feedback from $(\mb{y},\mb{y}_1)$ to  $\mb{x}$ and $\mb{x}_1$ is given by \cite[Theorem 17.3]{ElGamal:2012}
\begin{align}
C = \max_{\pdf(\mb{x},\mb{x}_1)} \min \Bset{\mb{I}(\mb{x},\mb{x}_1;\mb{y}) , \mb{I}(\mb{x};\mb{y},\mb{y}_1|\mb{x}_1) }
\label{capacity}
\end{align}
where $\mb{I}(.)$ is the mutual information function. 
%
For a \gls{hd} relay node we assume a \gls{td} access over the relay node as suggested in \cite{Host:2005} where the relay node only listens  in $\alfaA$ portion of the time ($0\leq \alfaA \leq 1$) and  transmits in the remaining $\overline{\alfaA} = 1-\alfaA$ portion. This results in the following
\begin{subequations}
\begin{align}\label{capacity-terms.1}
C_\mn{\tr{HD-1}} & = \mb{I}(\mb{x},\mb{x}_1;\mb{y}) = \alfaA \mb{I}(\mb{x};\mb{y}) + \overline{\alfaA} \mb{I}(\mb{x},\mb{x}_1;\mb{y}),\\
C_\mn{\tr{HD-2}} &  = \mb{I}(\mb{x};\mb{y},\mb{y}_1|\mb{x}_1) = \alfaA \mb{I}(\mb{x};\mb{y},\mb{y}_1) + \overline{\alfaA} \mb{I}(\mb{x};\mb{y}|\mb{x}_1),
\label{capacity-terms.2}
\end{align}
\end{subequations}
and the half-duplex capacity is
\begin{align}
C_\mn{\tr{HD}} = \max_{\pdf(\mb{x},\mb{x}_1)} \; \min \; \Bset{C_\mn{\tr{HD-1}},C_\mn{\tr{HD-2}} }.
\label{capacity.HD}
\end{align}
The source node can allocate a fraction $\alfaC$   of its energy ($0 \leq \alfaC \leq 1$) in the first  portion of time ($\alfaA$) and the remaining fraction $\overline{\alfaC} = 1 - \alfaC$  in the remaining portion $\overline{\alfaA}$. Therefore, for the \gls{awgn} channel \cite{ElGamal:2011}, the half-duplex capacity becomes
\begin{align}
C_\mn{\tr{HD}} = \max_{\alfaB, \alfaA, \alfaC} \; \min \; \Bset{C^\mn{\tr{AWGN}}_\mn{\tr{HD-1}},C^\mn{\tr{AWGN}}_\mn{\tr{HD-2}} },
\label{capacity.HD}
\end{align}
where
\begin{subequations}
\begin{align}\label{capacity-terms-3}
C^\mn{\tr{AWGN}}_\mn{\tr{HD-1}} & = \alfaA C\big(\frac{\alfaC}{\alfaA}(\gamma^\mc{SR} + \gamma^\mc{SD})\big) + \overline{\alfaA} C\big(\overline{\alfaB}\gamma^\mc{SR} \frac{\overline{\alfaC}}{\overline{\alfaA}}\big),
\\
C^\mn{\tr{AWGN}}_\mn{\tr{HD-2}} &=   \overline{\alfaA} C\left(\frac{\overline{\alfaC}}{\overline{\alfaA}} \gamma^\mc{SD}+\frac{1}{\overline{\alfaA}}\gamma^\mc{RD}+2\sqrt{\alfaB \frac{\overline{\alfaC}}{(\overline{\alfaA})^2}\gamma^\mc{SD}\gamma^\mc{RD}}\right)  + \alfaA C\Big(\frac{\alfaC}{\alfaA} \gamma^\mc{SD}\Big),
\label{capacity-terms-2}
\end{align}
\end{subequations}
with the ergodic form of $C_\mn{\tr{HD-erg}} = \ms{E}\{C_\mn{\tr{HD}}\}$. We can relax  $\alfaC$ parameter in the maximization in \eqref{capacity.HD}, for the sake of fixed-power transmission assumption, by choosing $\alfaC = \alfaA$ in \eqref{capacity-terms-2}. The particular case where only one transmitter node can be active at a time, is found by puting $\alfaB = 0$.

\section{Remarks on  Complexity of the Optimizations}
\label{Sec:Complexity}

%
In general, there is no  analytical formulas for the solution of a convex optimization problem however, there are  effective methods like the \emph{interior-point} methods that in some cases can  provably  solve the problem to a specified accuracy \cite{Boyd:2009}.



Here, we want to use a convex programming optimization method to solve the rate allocation  problem. The question is: Can we get a better solution  by locally optimizing the original problem and using the solution of the approximate problem $\tilde{\pi}$ as the starting point?

To answer this question we run a set of experiments using the ``\verb"fminsearch"'' function in MATLAB which is an interior-point optimization function. The experiments are on  optimizing  the original rate allocation problem in \eqref{TH.max}, using different starting points, as follows:
\begin{enumerate}
\item    Set the starting point at $0.1$ for all the optimization parameters (i.e., the redundancy values). We denote the result of this experiment by $\pi_\tr{o}$.
\item   Optimization using  $\tilde{\pi}$ (i.e., the solution to the two-dimensional approximated version of the problem) as the starting point. We denote the result of this experiment by $\tilde{\pi}_\tr{o}$ (or the optimized $\tilde{\pi}$).
\item   Starting point being set at  $\check{\pi}$ (i.e., the solution to the one-dimensional approximated version of the problem) with  the result of this experiment being denoted by $\check{\pi}_\tr{o}$.
\end{enumerate}

\begin{figure}[t!]
\psfrag{a}[c][c][\scalevalueS]{$\mc{S}$}
\psfrag{b}[c][c][\scalevalueS]{$\mc{R}$}
\psfrag{c}[c][c][\scalevalueS]{$\mc{D}$}
\psfrag{1}[c][c][\scalevalueS]{Source}
\psfrag{2}[c][c][\scalevalueS]{Relay}
\psfrag{3}[c][c][\scalevalueS]{Destination}
\psfrag{x}[c][c][\scalevalueS]{$d$}
\psfrag{y}[c][c][\scalevalueS]{$1-d$}
\begin{center}
\scalebox{1}{\includegraphics[width=\widthRatio\linewidth]{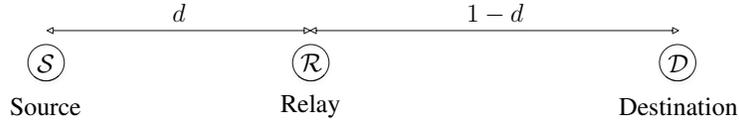}}
\caption{Topology of the relay channel under experiment.}\label{Fig:Topology.d}
\end{center}
\end{figure}

We run the tests for  the channel characteristics as follows. We  assume Rayleigh-fading links between the nodes. For a Rayleigh fading channel, the  \gls{snr} is characterized by the exponential \gls{pdf} of
\begin{align}\label{Pdf.Gamma}
p_{\gamma^\mf{ab}} = \frac{1}{\overline{\gamma}^\mf{ab}}\exp\bigg(-\frac{\gamma^\mf{ab}}{\overline{\gamma}^\mf{ab}}\bigg),
\end{align}
where $\overline{\gamma}^\mf{ab}$ is the average \gls{snr}. We also assume a channel with normalized distance of one between source and destination, and a relay node positioned with a distance of $0\leq d \leq 1$ from source on the line between source and relay as depicted in \figref{Fig:Topology.d}.

Therefore, the relation of the average long-term channel gain of the links between the nodes will be
\begin{subequations}
\label{Eq:gammaInSRandRD}
\begin{align}
\overline{\gamma}^\mn{SR} & = \frac{1}{d^\nu}\overline{\gamma}^\mn{SD}
\\ \overline{\gamma}^\mn{RD} & = \frac{1}{(1-d)^\nu}\overline{\gamma}^\mn{SD},
\end{align}
\end{subequations}
with $\nu$ being path-loss exponent. Unless otherwise specified, for all the numerical results in this paper we assume that $d = 0.5$ and we set the path-loss exponent $\nu = 4$.

\begin{figure}[tb]
\begin{center}
\psfrag{xlabel}[ct][ct][\scalevalueS]{$\overline{\gamma}^\mn{SD}$ [dB]}
\psfrag{ylabel}[c][c][\scalevalueS]{throughput, $\eta$}
\psfrag{XXXXXX1}[lc][lc][\scalevalueS]{$\eta(\tilde{\pi})$; $K=2$}
\psfrag{x2}[lc][lc][\scalevalueS]{$\eta(\tilde{\pi})$; $K=3$}
\psfrag{x3}[lc][lc][\scalevalueS]{$\eta(\tilde{\pi})$; $K=4$}
\psfrag{x4}[lc][lc][\scalevalueS]{$\eta(\tilde{\pi})$; $K=8$}
\psfrag{x5}[lc][lc][\scalevalueS]{$\eta(\tilde{\pi}_\tr{o})$; $K=4$}
\psfrag{x6}[lc][lc][\scalevalueS]{$\eta(\tilde{\pi}_\tr{o})$; $K=8$}
\psfrag{x7}[lc][lc][\scalevalueS]{$\eta(\tilde{\pi}_\tr{o})$; $K=4$}
\psfrag{x8}[lc][lc][\scalevalueS]{$\eta(\tilde{\pi}_\tr{o})$; $K=8$}
\psfrag{XXXXXx15}[lc][lc][\scalevalueS]{$\tilde{\eta}(\pi_\tr{o})$; $K=2$}
\psfrag{x16}[lc][lc][\scalevalueS]{$\tilde{\eta}_(\pi_\tr{o})$; $K=3$}
\psfrag{x17}[lc][lc][\scalevalueS]{$\tilde{\eta}(\pi_\tr{o})$; $K=4$}
\psfrag{x18}[lc][lc][\scalevalueS]{$\tilde{\eta}(\pi_\tr{o})$; $K=8$}
\psfrag{XXXXXx19}[lc][lc][\scalevalueS]{${\eta}(\check{\pi}_\tr{o})$; $K=2$}
\psfrag{x20}[lc][lc][\scalevalueS]{${\eta}(\check{\pi}_\tr{o})$; $K=3$}
\psfrag{x21}[lc][lc][\scalevalueS]{${\eta}(\check{\pi}_\tr{o})$; $K=4$}
\psfrag{x22}[lc][lc][\scalevalueS]{${\eta}(\check{\pi}_\tr{o})$; $K=8$}
\includegraphics[width=\widthRatio\linewidth,keepaspectratio]{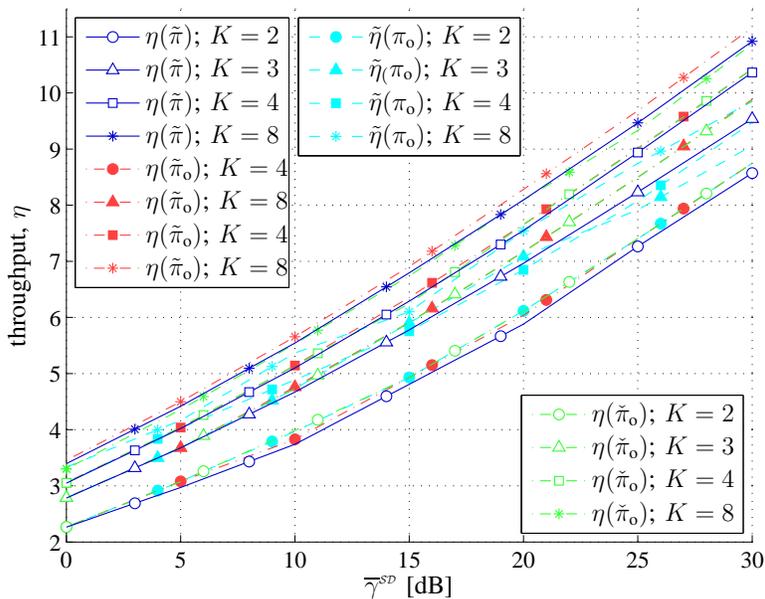}
\caption{Throughput for different optimization experiments.}\label{Fig:FSTH}
\end{center}
\end{figure}

\begin{figure}[tb!]
\begin{center}
\psfrag{xlabela}[lc][lc][\scalevalueS]{}
\psfrag{xlabelb}[ct][ct][\scalevalueS]{throughput, $\eta$}
\psfrag{ylabela}[c][c][\scalevalueS]{frequency}
\psfrag{ylabelb}[c][c][\scalevalueS]{frequency}
\psfrag{XXXXXXX1}[lc][lc][\scalevalueS]{freq. of $\eta(\pi_{r})$}
\psfrag{x4}[lc][lc][\scalevalueS]{$\eta(\tilde{\pi}_\tr{o})$}
\psfrag{x2}[lc][lc][\scalevalueS]{$\eta(\tilde{\pi})$}
\psfrag{x3}[lc][lc][\scalevalueS]{$\max \; \eta(\pi_{r})$}
\psfrag{XXXXXXX6}[lc][lc][\scalevalueS]{freq. of $\eta(\pi_{r})$}
\psfrag{x9}[lc][lc][\scalevalueS]{$\eta(\tilde{\pi}_\tr{o})$}
\psfrag{x7}[lc][lc][\scalevalueS]{$\eta(\tilde{\pi})$}
\psfrag{x8}[lc][lc][\scalevalueS]{$\max \; \eta(\pi_{r})$}
\psfrag{xa}[lc][lc][\scalevalueS]{(a)}
\psfrag{xb}[lc][lc][\scalevalueS]{(b)}
\includegraphics[width=\widthRatio\linewidth,keepaspectratio]{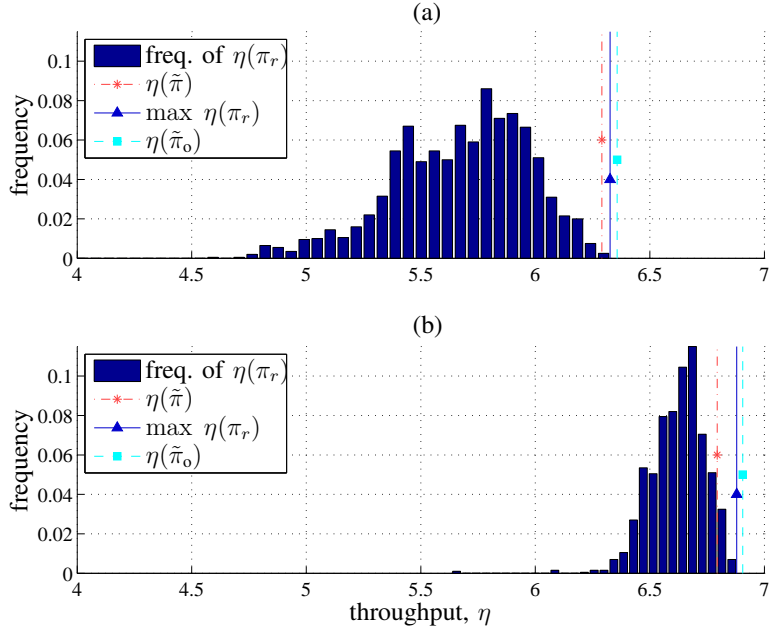}
\caption[Foo Bar.]{Histogram plot for the normalized frequency of a solution $\pi_{r}$ that is found using the MATLAB optimization function with randomly generated starting point for $\overline{\gamma}^\mn{SD} = 15$ dB for: (a) $K=4$ and (b) $K = 8$. The maximum throughput found in the random starting point experiment and the maximum throughput found using the proposed optimization method ($\eta(\tilde{\pi})$) is shown for comparison.}\label{Fig:FSP}
\end{center}
\end{figure}

The results of the maximum achieved throughput with each of the above experiments are shown in \figref{Fig:FSTH}. The optimization experiments result in a slightly improved throughput value in all the cases except for the first experiment where a random point is given to the optimization algorithm as an starting point. This magnifies the importance of the starting point in a non-linear optimization problem.

Experiment results for the second test that we run are shown in \figref{Fig:FSP}.  In this test we try to globally optimize the throughput using randomly generated starting points $\pi_\tr{r}$. We repeated the test for 2000 randomly generated starting points. For $K = 4$,  $\sim 96$\% of the tests converged to a solution with the  values depicted in \figref{Fig:FSP}, while only 0.15 \% of the results are in the range of $\eta(\tilde{\pi})$ or larger. For $K = 8$ the convergence rate is only $\sim$65 \%.

The optimal  result of this test is less than the result of optimization result  when the starting point is set to $\tilde{\pi}$ which is shown as $\tilde{\pi}_\tr{o}$ in \figref{Fig:FSP}. This is despite the fact that  finding $\tilde{\pi}$ and then  $\tilde{\pi}_\tr{o}$  takes at most a few hours of time on a regular personal computer for $K = 8$ while the random starting point test above takes time in order of weeks on the same computer. 

\section{Numerical Results}
\label{Sec:NumResults}

We use the 2-D state-space  solution for maximum throughput in \secref{Ap:DP} to derive the solutions denoted by $\tilde{\pi}$. The  $\tilde{\pi}$ solution is then  optimized according to the third experiment explained in \secref{Sec:Complexity} in order to find $\tilde{\pi}_\tr{o}$, as the maximum throughput achieving variable rate policy. The results in this section are derived for the same Rayleigh block fading channel described  in \secref{Sec:Complexity}, where the average long-term channel gains follow from \eqref{Eq:gammaInSRandRD}.

\begin{figure}[h]
\begin{center}
\psfrag{xlabel}[lt][lb][\scalevalueS]{$k$}
\psfrag{ylabel}[c][c][\scalevalueS]{redundancy, $\rho$}
\psfrag{t1}[lc][lc][\scalevalueS]{${\rho}^\mn{S}_k$}
\psfrag{t2}[lc][lc][\scalevalueS]{${\rho}^\mn{R}_{1,k}$}
\psfrag{t3}[lc][lc][\scalevalueS]{${\rho}^\mn{R}_{2,k}$}
\psfrag{t4}[lc][lc][\scalevalueS]{${\rho}^\mn{R}_{3,k}$}
\psfrag{XX1}[lc][lc][\scalevalueS]{$\tilde{\pi}_\tr{o}$}
\psfrag{x2}[lc][lc][\scalevalueS]{$\check{\pi}$}
\psfrag{x3}[lc][lc][\scalevalueS]{$\tilde{\pi}$}
\psfrag{x4}[lc][lc][\scalevalueS]{$\hat{\rho}$; FR}
\includegraphics[width=\widthRatio\linewidth,keepaspectratio]{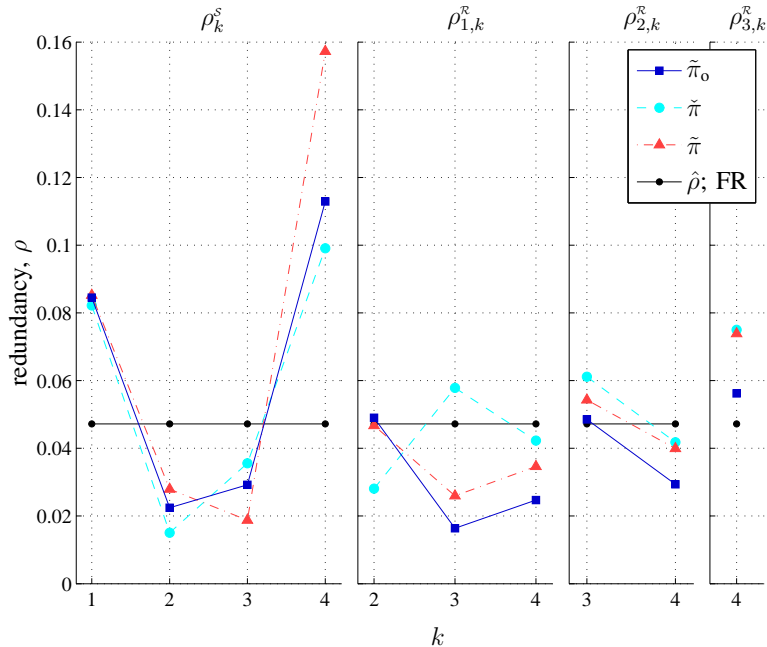}
\caption[Foo Bar.]{Optimal policy $\tilde{\pi}$ found with the two dimensional optimization method and $\check{\pi}$ from the one-dimensional simplified method  for $\overline{\gamma}^\mn{SD} = 15$ dB and $K = 4$, compared to the optimal fixed-rate (FR) redundancy value $\hat{\rho}$.}
\label{Fig:OptRho}
\end{center}
\vspace{-10pt}
\end{figure}

Examples of the optimized variable-rate HARQ transmission policies are shown in \figref{Fig:OptRho}.  The maximum throughput achieving set $\tilde{\pi}_\tr{o}$ is compared with  $\tilde{\pi}$ as the result of the  proposed 2-D state-space solution and $\check{\pi}$ which is found using the proposed 1-D state-space solution. As shown in \figref{Fig:OptRho} for the \gls{snr} of $\overline{\gamma}^\mn{SD} = 15$ dB, the three presented solutions seem to follow the same trend with $\tilde{\pi}_\tr{o}$ policy showing a close solution to $\tilde{\pi}$ as expected. An optimal throughput achieving encoder has a decision making process according to the solution  result in \figref{Fig:OptRho} as follows: $\mc{S}$ starts the transmission process by choosing a subset of $N^\mn{S}_{\tr{s},1} = \rho^\mn{S}_1 \cdot N_\tr{b}$ number of symbols from the generated codeword $\mb{x}$ and broadcasts it to the other two nodes. Retransmissions from node $\mc{S}$ will then be pursued using  $N^\mn{S}_{\tr{s},2} = \rho^\mn{S}_2 \cdot N_\tr{b}$ new symbols from the same codeword. In the case that $\mc{R}$ successfully  decodes the message, then the encoder in node $\mc{R}$ will create a subcodeword with length  $N^\mn{R}_{\tr{s},2} = \rho^\mn{R}_{1,2} \cdot N_\tr{b}$ and takes over the \gls{harq} transmission. This process will continue until $\mc{D}$ decodes the message successfully or a maximum $K = 4$ transmissions is achieved.

\begin{figure}[]
\begin{center}
\psfrag{xlabel}[lt][lb][\scalevalueS]{$\overline{\gamma}^\mn{SD}$ [dB]}
\psfrag{ylabel}[c][c][\scalevalueS]{throughput, $\eta$}
\psfrag{XXXXX1}[lc][lc][\scalevalueS]{$K = 2$; FR}
\psfrag{x2}[lc][lc][\scalevalueS]{$K = 3$; FR}
\psfrag{x3}[lc][lc][\scalevalueS]{$K = 4$; FR}
\psfrag{x0}[lc][lc][\scalevalueS]{$K = 8$; FR}
\psfrag{XXXXXx20}[lc][lc][\scalevalueS]{$\eta_{\mbox{\tiny{max}}}; K=\infty$}
\psfrag{x21}[lc][lc][\scalevalueS]{$\eta_{\mbox{\tiny{0}}}; K=1$}
\psfrag{x22}[lc][lc][\scalevalueS]{$C_\mn{\tr{HD-erg}}$}
\psfrag{x23}[lc][lc][\scalevalueS]{$C_\mn{\tr{HD-erg}}$; $\beta = 0$}
\psfrag{XXXXX5}[lc][lc][\scalevalueS]{$K = 2$; VR}
\psfrag{x16}[lc][lc][\scalevalueS]{$K = 3$; VR}
\psfrag{x17}[lc][lc][\scalevalueS]{$K = 4$; VR}
\psfrag{x18}[lc][lc][\scalevalueS]{$K = 8$; VR}
\includegraphics[width=\widthRatio\linewidth,keepaspectratio]{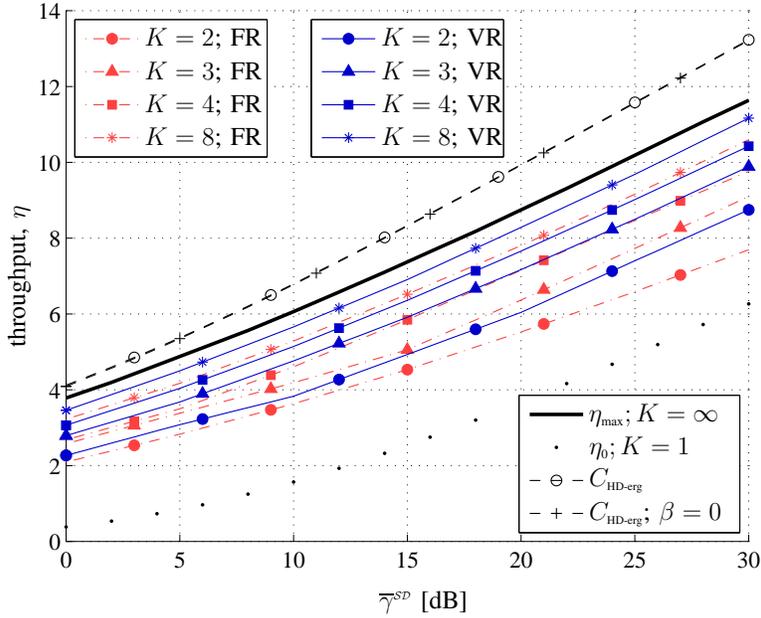}
\caption{Maximum achieved throughput for the proposed variable-rate (VR) scheme $\eta(\tilde{\pi}_\tr{o})$ compared to the maximum achievable throughput for fixed-rate (FR) transmission. The performance limit for the described channel is also shown for $K = \infty$ of adaptive-rate HARQ transmission. The obvious lower bound
of one transmission for $K = 1$ (the direct transmission lower bound) is also shown in the figure.}
\label{Fig:OptTH}
\end{center}
\vspace{-10pt}
\end{figure}

In \figref{Fig:OptTH} the maximum achieved throughput using the variable-rate (VR) transmission method proposed in this paper is compared to the maximum throughput achievable for fixed-rate (FR) transmission. The maximum achievable throughput for $K = \infty$ of adaptive-rate transmission as explained by the authors in  \cite{Khosravirad:2013} is shown in \figref{Fig:OptTH} along with the truncated \gls{harq} results. We can see that using   the sub-optimal variable-rate transmission method presented in this paper we can get as close as a 2 dB  to the maximum achievable throughput with $K = 8$, where for the fixed-rate transmission this difference is  4-5 dB. As already discussed in \cite{Leszek:2013}, the performance of variable-rate \gls{harq} is upper-bounded by adaptive-rate \gls{harq}. However, in some cases costs of extra feedback bits can become too high  for the communication network and a single-bit \gls{ack}/\gls{nack} can only be provided to the link. For example, in the \gls{lte} \gls{ul} control channel, it will cost 1 ms of the resources for  a low-coverage user to transmit a single bit feedback message \cite{3gpp36912} which will make it highly impractical to  schedule such a user with more than one bit in the \gls{ul} as feedback message. Therefore, extra information about the state of the decoder is out of budget for this type of users and the variable-rate \gls{harq} transmission will be the more practical choice.

\begin{figure}[]
\begin{center}
\psfrag{xlabel}[lt][lb][\scalevalueS]{$\overline{\gamma}^\mn{SD}$ [dB]}
\psfrag{ylabel}[c][c][\scalevalueS]{outage probability, $\Pout$}
\psfrag{XXXXX1}[lc][lc][\scalevalueS]{$K = 2$; VR}
\psfrag{x2}[lc][lc][\scalevalueS]{$K = 3$; VR}
\psfrag{x3}[lc][lc][\scalevalueS]{$K = 4$; VR}
\psfrag{XXXXX7}[lc][lc][\scalevalueS]{$K = 2$; FR}
\psfrag{x8}[lc][lc][\scalevalueS]{$K = 3$; FR}
\psfrag{x9}[lc][lc][\scalevalueS]{$K = 4$; FR}
\includegraphics[width=\widthRatio\linewidth,keepaspectratio]{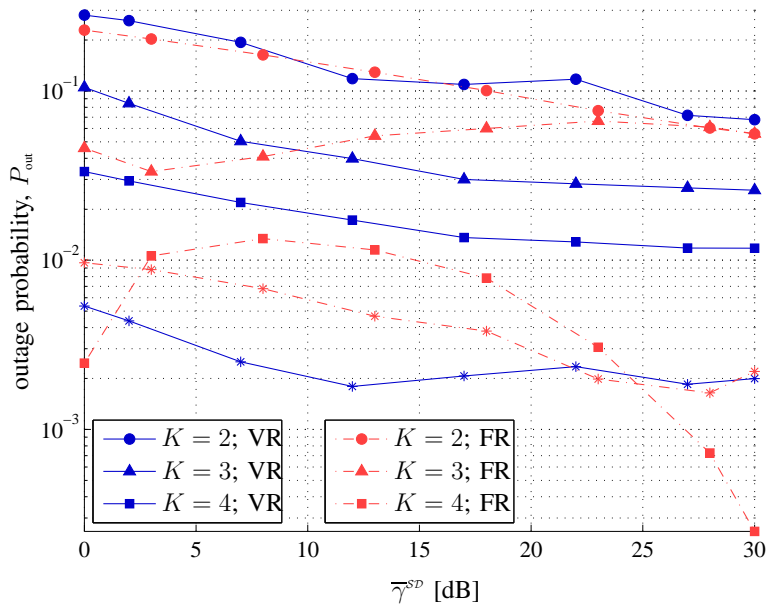}
\caption{Outage probability for the optimal throughput achieving $\tilde{\pi}_\tr{o}$.}
\label{Fig:OptOut}
\end{center}
\vspace{-10pt}
\end{figure}

The presented results in \figref{Fig:OptTH} shows that increasing  $K$ can significantly improve the throughput performance of the system model in order to reach the maximum achievable throughput $\eta_{\mbox{\tiny{max}}}$. This however will increase the average delivery time, it will result also in lower outage probability as shown in \figref{Fig:OptOut}. As depicted, the optimal throughput approach tends to keep a consistent outage probability  $\Pout$ for different average \gls{snr} values. The decreasing trend of the outage probability with respect to $K$ confirms once again that the  a capacity approaching  high throughput with arbitrarily low outage probability can be reached in the block-fading channel by choosing a large enough $K$ maximum transmissions. Finally, for the results in \figref{Fig:OptTH} and \figref{Fig:OptOut} for different \gls{snr} values and for $K = 2, 3, 4, 8$, we can see a significant improvement on the average throughput for the proposed variable-rate method compared to the fixed-rate transmission. the fixed-rate transmission results according to \figref{Fig:OptOut} can reach the same outage probability as the proposed method only at the cost of losing average throughput.

In \figref{Fig:RelPos}, we study the effect of relay position parameter $d$ on the maximum achievable throughput. For the proposed variable-rate method, the maximum  throughput is achieved  at $d = 0.5$. For the fixed-rate transmission though, especially for $K = 2, 3$ maximum throughput happens when the relay node is closer to the source node.

\section{Conclusions}\label{Sec:Conclusions}

We analyzed a variable-rate incremental redundancy \gls{harq} transmission for relay-based cooperative  transmission. The main difficulty of optimizing the transmission rates was  addressed via doubly-recursive \gls{dp}, using suitable approximations of the outage probability. The numerical results obtained in various topologies show that the proposed variable-rate cooperative \gls{harq} scheme (i) significantly improves the throughput compared to the fixed-rate counterpart, (ii) is comparable  to the \gls{csi}-aware relaying for relatively low \gls{snr}, and (iii) looses with respect to \gls{csi}-aware solution for high \gls{snr}.


\begin{appendices}

\begin{figure}[t]
\begin{center}
\psfrag{xlabel}[lt][lb][\scalevalueS]{$d$}
\psfrag{ylabel}[lc][lc][\scalevalueS]{throughput, $\eta$}
\psfrag{XXXXXx1}[lc][lc][\scalevalueS]{$K = 2$, VR}
\psfrag{X2}[lc][lc][\scalevalueS]{$K = 3$, VR}
\psfrag{X3}[lc][lc][\scalevalueS]{$K = 4$, VR}
\psfrag{X4}[lc][lc][\scalevalueS]{$K = 8$, VR}
\psfrag{XXXXx5}[lc][lc][\scalevalueS]{$K = 2$, FR}
\psfrag{X6}[lc][lc][\scalevalueS]{$K = 3$, FR}
\psfrag{X7}[lc][lc][\scalevalueS]{$K = 4$, FR}
\psfrag{X8}[lc][lc][\scalevalueS]{$K = 8$, FR}
\psfrag{XXXXxx8}[lc][lc][\scalevalueS]{$\eta_{\mbox{\tiny{max}}}, K=\infty$}
\psfrag{X9}[lc][lc][\scalevalueS]{$\eta_{\mbox{\tiny{0}}}, K=1$}
\psfrag{X10}[lc][lc][\scalevalueS]{$C_\mn{\tr{HD-erg}}$}
\psfrag{X11}[lc][lc][\scalevalueS]{$C_\mn{\tr{HD-erg}}$; $\beta = 0$}
\includegraphics[width=\widthRatio\linewidth,keepaspectratio]{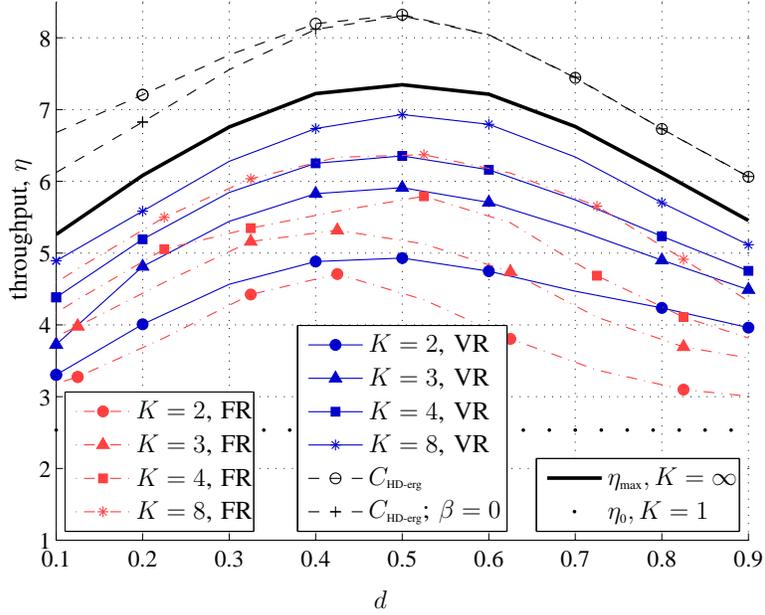}
\caption[Foo Bar.]{Maximum achieved throughput for the proposed variable-rate (VR) scheme $\eta(\tilde{\pi}_\tr{o})$ with respect to $d$, the distance of the relay node from the source node compared to the maximum achievable throughput for fixed-rate (FR) and the performance limit for the described channel for $\overline{\gamma}^\mn{SD} = 15$ dB.}
\label{Fig:RelPos}
\end{center}
\vspace{-25pt}
\end{figure}

\section{Throughput of Cooperative Variable-Rate HARQ Transmission}
\label{Ap:ThCalcAppendix}

A failure happens in the truncated HARQ process only if after $K$ transmission rounds $I_K^{\mn{D}} < 1$. This can result from  $K$ disjoint events given in \eqref{Eq:fail.l} and \eqref{Eq:fail.K}.
\begin{figure*}[t]
\begin{align}\label{Eq:fail.l}
E^{\ast}_{l}  = \Bset{  \sum_{k = 1}^{l-1} \iota_k^{\mn{SR}} < 1  \; \wedge \; \sum_{k = 1}^{l} \iota_k^\mn{SR} > 1   \; \wedge \; \sum_{k = 1}^{l}\iota_k^{\mn{SD}} + \sum_{k = l+1}^{K}\iota_{l,k}^\mn{RD} < 1 },  \quad  1\leq l \leq K-1
\end{align}
\vspace*{-15pt}
\end{figure*}
\begin{align}\label{Eq:fail.K}
E^{\ast}_{K}  = \bset{ \sum_{k = 1}^{K-1} \iota_k^\mn{SR} < 1  \; \wedge \; \sum_{k = 1}^{K}\iota_k^\mn{SD}< 1 }.
\end{align}

There are also success events which we categorize into two groups. Some of the success events happen in the broadcasting phase  which we denote by $E_{l}$ and the other success events happen following a transition to the relaying phase at transmission attempt $l$ and we denote them by $E_{l,k}$. In other words, in the first group of these events, decoding at $\mc{D}$ is done only based on the information from  node $\mc{S}$ while in the second group, $\mc{R}$ has succeeded in decoding at some time $l$ and therefore, the destination node has some mutual information from the relay node too. These events can be presented in \eqref{Eq:Success.k} and \eqref{Eq:Success.l.k}.
\begin{figure*}[t]
\begin{align}
E_{l} &=  \Bset{  \sum_{i=1}^{l-1} \iota_i^\mn{SD}<1 \; \wedge  \; \sum_{i=1}^{l-1} \iota_i^\mn{SR}<1 \; \wedge \; \sum_{i=1}^{l} \iota_i^\mn{SD} >1  },  \quad  1\leq l \leq K
\label{Eq:Success.k}\\
E_{l,k} &=  \Bset{  \sum_{i=1}^{l-1} \iota_i^\mn{SR}<1 \; \wedge \; \sum_{i=1}^{l} \iota_i^\mn{SD} + \sum_{i=l+1}^{k-1} \iota_{l,i}^\mn{RD} <1\;  \wedge \; \sum_{i=1}^{l} \iota_i^\mn{SR}>1  \; \wedge \;  \sum_{i=1}^{l} \iota_i^\mn{SD} +\sum_{i=l+1}^{k} \iota_{l,i}^\mn{RD} >1   }, \nonumber \\ &  \quad  1\leq l < k \leq K
\label{Eq:Success.l.k}
\end{align}
\vspace*{-20pt}
\end{figure*}


From \corref{Corollary}, all the success events in \eqref{Eq:Success.k} and \eqref{Eq:Success.l.k} are mutually exclusive. The same way we can show that the success events and the failure events in \eqref{Eq:fail.l} and \eqref{Eq:fail.K} are disjoint too.

Probability of a failure event $E^{\ast}_{k}$ can be  represented using \eqref{Eq:P-one}--\eqref{P-three}. For instance from \eqref{Eq:fail.K} we have the probability of event $E^{\ast}_{K}$ as follows:
\begin{align}
\Pr\{E^{\ast}_{K}\}  =  \Pr \bset{ \sum_{k = 1}^{K-1} \iota_k^\mn{SR} < 1} \cdot \Pr \bset{ \sum_{k = 1}^{K}\iota_k^\mn{SD}< 1 }  =
\po{K} \cdot \pt{K-1}
\label{Eq:prob-EK}
\end{align}

For any two random events $A$ and be $B$ we know that $P(A\cap B)=P(A)-P(A\cap B^c)$, where $B^c$ is the complement the event $B$ which  gives us
{\fontsize{10.3pt}{10pt}\selectfont
\begin{align}
\Pr\{E^{\ast}_{l}\}  = \Pr \bset{ \sum_{i = 1}^{l-1} \iota_i^\mn{SR}<1 \; \wedge \; \sum_{i=1}^{l} \iota_i^\mn{SD} + \sum_{i = l+1}^{K} \iota_{l,i}^\mn{RD} < 1 }  -  \Pr \bset{ \sum_{i = 1}^{l} \iota_i^\mn{SR}<1 \;  \wedge \; \sum_{i=1}^{l} \iota_i^\mn{SD} + \sum_{i = l+1}^{K} \iota_{l,i}^\mn{RD} < 1  },
\label{Eq:Prob-Ek-1}
\end{align}}
which results in the following:
\begin{align}
\Pr \{E^{\ast}_{l}\} = \big[\pt{l-1}-\pt{l}\big]\cdot \ptr{l}{K}
\label{Eq:Prob-Ek-2}
\end{align}
The same way, we can find the probability of the success events as follows:
\begin{subequations}
\begin{align}\label{Eq:Prob-El-success}
\Pr \{E_{k}\} & = \big[\po{k-1}-\po{k}\big]\cdot \pt{k-1} \\
\Pr \{E_{l,k}\} & = \left\{\begin{matrix}
\big[\pt{l-1}-\pt{l}\big]\cdot \big[ \po{k-1} - \ptr{l}{k} \big] &  k = l+1 \\
\big[\pt{l-1}-\pt{l}\big]\cdot \big[ \ptr{l}{k-1} - \ptr{l}{k} \big] & k > l + 1
\end{matrix}\right.
\label{Eq:Prob-Elk-success}
\end{align}
\end{subequations}

An outage in message delivery in the transmission process can happen due to any of the failure events $E^{\ast}_{1}, \cdots, E^{\ast}_{K}$. Therefore, the outage probability can be shown as follows:
\begin{equation}\label{Eq:Pout.AL-M.0}
\Pout = \Pr \{  \cup_{k=1}^K E^{\ast}_{k} \}.
\end{equation}
Because the failure events  are mutually exclusive, \eqref{Eq:Pout.AL-M.0} can be shown as follows:
\begin{equation}\label{Eq:Pout.AL-M.1}
\Pout = \sum_{k=1}^K \Pr\{E^{\ast}_{k}\}.
\end{equation}
Substituting \eqref{Eq:prob-EK} and \eqref{Eq:Prob-Ek-2} in \eqref{Eq:Pout.AL-M.1} gives us \eqref{Eq:Pout-ALM}.

The expected number of channel uses $\overline{N}_\tr{s}$ in \eqref{Eq:Th}, is the expectation over the number of channel uses of all the possible events. Thus it can be shown as in \eqref{Eq:Ns-bar}
\begin{figure*}[t]
{\fontsize{11pt}{10pt}\selectfont
\begin{align}
\overline{N}_\tr{s} = N_\tr{b} \cdot \Big(& \sum_{k=1}^{K}\Pr\{E_{k}\}\cdot q_k  +  \sum_{l=1}^{K-1} \sum_{k=l+1}^{K} \Pr\{E_{l,k} \}\cdot q_{l,k} + \sum_{k=1}^{K-1}\Pr\{E^{\ast}_{k}\} \cdot q_{k,K} + \Pr\{E^{\ast}_{K}\} \cdot q_{K} \Big)
\label{Eq:Ns-bar}
\end{align}}
\vspace*{-30pt}
\end{figure*}
where, $q_k = \sum_{i=1}^{k}\rho_k^\mn{S}$ and $q_{l,k} = \sum_{i=1}^{l}\rho_i^\mn{S} + \sum_{i=l+1}^{k}\rho_{l,i}^\mn{R}$. Substituting \eqref{Eq:prob-EK}, \eqref{Eq:Prob-Ek-2}, \eqref{Eq:Prob-Elk-success} and \eqref{Eq:Prob-Elk-success} in \eqref{Eq:Ns-bar} gives us \eqref{Eq:Den-ALM}.

One can easily investigate the fact that all success and failure events create a set of disjoint events where the sum of their probabilities equals 1. This is shown in the following.
\begin{align}
\sum_{k=1}^{K}\left(\Pr\{E_{k}\} + \Pr\{E^{\ast}_{k}\} \right) + \sum_{l=1}^{K-1} \sum_{k=l+1}^{K} \Pr\{E_{l,k} \} = 1.
\label{Eq:ProbCheck}
\end{align}


\section{Throughput of $M$-Relay Network}
\label{Ap:MThCalcAppendix}

We denote the event of going through path $l$ by $E(l)$. For failure paths, the event $E(l)$ can be shown  as in \eqref{EQ:Ap111}.
\begin{figure*}[t]
\begin{align}\nonumber
E(l) =  \{ & I_1^{\mnR_\mnn{l_1}\mnD}< 1 \;  \wedge \; I_1^{\mnR_\mnn{l_1}\mnR_\mnn{l_2}} > 1 \;  \wedge \;  I_1^{\mnR_\mnn{l_1}\mnR_\mnn{j}}< 1 |_{\forall j>l_2} \; \wedge \\\nonumber
& I_1^{\mnR_\mnn{l_1}\mnD}+ I_2^{\mnR_\mnn{l_2}\mnD}< 1 \; \wedge \; I_1^{\mnR_\mnn{l_1}\mnR_\mnn{l_3}} + I_2^{\mnR_\mnn{l_2}\mnR_\mnn{l_3}} > 1 \; \wedge \;  I_1^{\mnR_\mnn{l_1}\mnR_\mnn{j}} + I_2^{\mnR_\mnn{l_2}\mnR_\mnn{j}}< 1 |_{\forall j>l_3} \; \wedge \\\nonumber
& ... \; . \; ... \; . \;  ... \; . \\
& \sum_{k =1}^{K} I_k^{\mnR_\mnn{l_{k}}\mnD}< 1 \; \wedge \;  \sum_{k =1}^{K} I_k^{\mnR_\mnn{l_{k}}\mnR_\mnn{l_\Kp}}  > 1 \; \wedge \; \sum_{k =1}^{K} I_k^{\mnR_\mnn{l_{k}}\mnR_\mnn{j}}< 1 |_{\forall j>l_\Kp}   \}  \label{EQ:Ap111}
\end{align}
\vspace*{-10pt}
\end{figure*}
Then we can  reduce \eqref{EQ:Ap111} into the following:
\begin{align}\nonumber
E(l)  =  \{ & \sum_{k =1}^{K} I_k^{\mnR_\mnn{l_{k}}\mnD}< 1 \; \wedge \; I_1^{\mnR_\mnn{l_1}\mnR_\mnn{l_2}} > 1 \;  \wedge  I_1^{\mnR_\mnn{l_1}\mnR_\mnn{l_3}} + I_2^{\mnR_\mnn{l_2}\mnR_\mnn{l_3}} > 1 \; \wedge \; I_1^{\mnR_\mnn{l_1}\mnR_\mnn{j}} + I_2^{\mnR_\mnn{l_2}\mnR_\mnn{j}}< 1 |_{l_{2} <j \leq l_{3}} \; \wedge \\ \nonumber
& \sum_{k =1}^{3} I_k^{\mnR_\mnn{l_{k}}\mnR_\mnn{l_4}}  > 1 \; \wedge \; \sum_{k =1}^{3} I_k^{\mnR_\mnn{l_{k}}\mnR_\mnn{j}}< 1 |_{l_{3} <j \leq l_{4}} \; \wedge \\ \nonumber
& ... \; . \; ... \; . \;  ... \; . \\ 
& \sum_{k =1}^{K} I_k^{\mnR_\mnn{l_{k}}\mnR_\mnn{l_\Kp}}  > 1 \; \wedge \; \sum_{k =1}^{K} I_k^{\mnR_\mnn{l_{k}}\mnR_\mnn{j}}< 1 |_{l_{K} <j \leq l_{\Kp}} \}  \label{EQ:Ap112}
\end{align}

Probability of  a failure event $E(l)$ in \eqref{EQ:Ap112}  can then be shown as in \eqref{EQ:Ap113}.
\begin{figure*}[t]
{\fontsize{10.3pt}{10pt}\selectfont
\begin{align}\nonumber
\PR{E(l)}  =   & \PR{ \sum_{k=1}^{K}I_k^{\mnR_\mnn{l_{k}}\mnD}< 1} \times  (1 - \PR{I_1^{\mnR_\mnn{l_1}\mnR_\mnn{l_2}} < 1}) \times \\ 
& \prod_{\theta = 2}^K \left ( \PR{\sum_{k =1}^{\theta-1} I_k^{\mnR_\mnn{l_{k}}\mnR_\mnn{l_{\theta+1}}} < 1} - \PR{\sum_{k =1}^{\theta} I_k^{\mnR_\mnn{l_{k}}\mnR_\mnn{l_{\theta+1}}} < 1} \right ) \times  \prod_{\theta = 2}^{K+1} \prod_{l_{\theta}<j<l_{\theta+1}} \PR{\sum_{k=1}^{\theta-1}I_k^{\mnR_\mnn{l_{k}}\mnR_\mnn{j}} < 1}  \label{EQ:Ap113}
\end{align}}
\vspace*{-20pt}
\end{figure*}
In a similar approach as for failure events, the event $E(l)$ of going  through a success path $l$, can  be shown as in \eqref{EQ:Ap114}.
\begin{figure*}[t]
{\fontsize{10.3pt}{10pt}\selectfont
\begin{align}\nonumber
E(l)  =  \{ & I_1^{\mnR_\mnn{l_1}\mnD}< 1 \; \wedge \; I_1^{\mnR_\mnn{l_1}\mnR_\mnn{l_2}} > 1 \; \wedge \;  I_1^{\mnR_\mnn{l_1}\mnR_{j:l_2<j\leq l_3}}< 1 \; \wedge \\ \nonumber & I_1^{\mnR_\mnn{l_1}\mnD}+ I_2^{\mnR_\mnn{l_2}\mnD}< 1 \; \wedge \; I_1^{\mnR_\mnn{l_1}\mnR_\mnn{l_3}} + I_2^{\mnR_\mnn{l_2}\mnR_\mnn{l_3}} > 1 \; \wedge \;  I_1^{\mnR_\mnn{l_1}\mnR_{j}} + I_2^{\mnR_\mnn{l_2}\mnR_{j:l_3<j\leq l_4}}< 1 \; \wedge \\\nonumber
& ... \; . \; ... \; . \;  ... \; . \\ 
&\sum_{i=1}^{k-1} I_i^{\mnR_\mnn{l_{k}}\mnD}< 1 \; \wedge \; \sum_{i = 1}^{k-1} I_{i}^{\mnR_\mnn{l_{i}}\mnR_\mnn{l_{k}}} > 1 \; \wedge  \;  I_1^{\mnR_\mnn{l_1}\mnR_{j}} + ... + I_{k-1}^{\mnR_\mnn{l_{k-1}}\mnR_{j:l_{k}<j<l_{k+1}}}< 1 \; \wedge  I_1^{\mnR_\mnn{l_1}\mnD}+ ...+ I_{k}^{\mnR_\mnn{l_{k}}\mnD}>1  \} \label{EQ:Ap114}
\end{align}}
\vspace*{-20pt}
\end{figure*}
The  probability of the success event   $E(l)$ is as in \eqref{EQ:Ap115}.
\begin{figure*}[t]
{\fontsize{10.3pt}{10pt}\selectfont
\begin{align}\nonumber
\PR{E(l)}  =   & \left ( \PR{ \sum_{i=1}^{k-1}I_i^{\mnR_\mnn{l_{i}}\mnD} < 1 } - \PR{ \sum_{i=1}^{k}I_i^{\mnR_\mnn{l_{i}}\mnD} < 1 } \right ) \times  (1 - \PR{I_1^{\mnR_\mnn{l_1}\mnR_\mnn{l_2}} < 1}) \times \\ 
&\prod_{\theta = 2}^{k-1} \left ( \PR{\sum_{k =1}^{\theta-1} I_k^{\mnR_\mnn{l_{k}}\mnR_\mnn{l_{\theta+1}}} < 1} - \PR{\sum_{k =1}^{\theta} I_k^{\mnR_\mnn{l_{k}}\mnR_\mnn{l_{\theta+1}}} < 1} \right ) \times \prod_{\theta = 2}^{k} \prod_{l_{\theta}<j<l_{\theta+1}} \PR{\sum_{k=1}^{\theta}I_k^{\mnR_\mnn{l_{k}}\mnR_\mnn{j}} < 1} \label{EQ:Ap115}
\end{align}}
\vspace*{-20pt}
\end{figure*}

%

With the definition in \eqref{Eq:PLOalpha}, we can find the probability of failure and success paths, respectively as follows.
\begin{align}
\PR{E(l)}  =  \plo{K}{\mcD} \times \prod_{\theta = 1}^K \left ( \plo{\theta-1}{\mcR_{l_{\theta+1}}} - \plo{\theta}{\mcR_{l_{\theta+1}}} \right )  \times  \prod_{\theta = 2}^{K+1} \prod_{l_{\theta}<j<l_{\theta+1}} \plo{\theta-1}{\mcR_{j}} \quad  l \in \tr{failure events}.
\end{align}\vspace{-10pt}
\begin{align}\nonumber
\PR{E(l)}  = & \left ( \plo{k-1}{\mcD}-\plo{k}{\mcD} \right )  \times \prod_{\theta = 1}^{k-1} \left (  \plo{\theta-1}{\mcR_{l_{\theta+1}}} - \plo{\theta}{\mcR_{l_{\theta+1}}} \right ) \times \\ & \prod_{\theta = 2}^{k} \prod_{l_{\theta}<j<l_{\theta+1}} \plo{\theta-1}{\mcR_{j}} \quad  l \in \tr{success events},
\end{align}
where, by definition we have  $\plo{0}{\alpha} = 1$.

An outage in message delivery in the transmission process can happen due to any of the failure events. Therefore, the outage probability can be shown as follows:
\begin{equation}
\Pout = \Pr \{  \cup_{l \in failure} E(l) \} = \sum_{l \in failure} \Pr\{E(l)\},
\end{equation}
which follows from the events being disjoint. Moreover, the denominator of the throughput in \eqref{Eq:Eta-ALMM} can be found as follows,
\begin{align}
\D  = \sum_{l} \Pr\{E(l)\} \cd \overline{\vrho}^l,
\label{Eq:Den-ALMM}
\end{align}
with $\overline{\vrho}^l$ from \eqref{Eq:vrhol}.


\section{One Dimensional State Space Optimization}

\label{Ap:OneDimensionalState}
\begin{figure*}[tb!]
\begin{subequations}
\begin{align}\label{Eq:Ulambdak}
U^{\lambda,i}_{k}(X,\xp) & = \min_{ \substack{\rho}} U^{\lambda,i}_{k-1}(X,\xp -\rho) + \rho \cdot \cptr(X,\xp -\rho),
\\\label{Eq:Ulambda2}
U^{\lambda,i}_{i+2}(X,\xp) & = \min_{ \substack{\rho}} (\xp -\rho)\cd \cpo(X) + \rho \cdot \cptr(X,\xp -\rho).
\end{align}
\end{subequations}
\vspace*{-20pt}
\end{figure*}
\begin{figure*}[tb!]
{\fontsize{10.3pt}{10pt}\selectfont
\begin{subequations}
\begin{align}\label{Eq:Jlambda.OneDim}
J^{\lambda}_K(X)  &= \min_{ \substack{\rho^{\mc{S}}_{1}, \ldots , \rho^{\mc{S}}_{K}  \\ \sum_{k=1}^{K} \rho^{\mc{S}}_{k} = X}} \Bset{ \sum_{i=1}^{K-1} \big[\rho^\mn{S}_i \cd \ttpo{i-1} \cd \ttpt{i-1}\big] + \check{f}_i \cd \check{U}^{\lambda,i}(X)
 + \lambda \cd \ttpo{K} \cd \ttpt{K-1} + \rho^\mn{S}_K \cd \ttpo{K-1} \cd \ttpt{K-1}}
\\ & = \min_{ \substack{\rho = \rho^{\mn{S}}_{K}}} J^{\lambda}_{K-1}(X-\rho) + \rho \cdot \cpo(X-\rho) \cd \cpt(X-\rho) + \lambda \cd \cpo(X)\cd \cpt(X-\rho) \label{Eq:Jlambda.OneDim.K}
\\\label{Eq:Jlambda.OneDim.k}
J^{\lambda}_{k}(X) & = \min_{ \substack{\rho}} J^{\lambda}_{k-1}(X -\rho) + \rho \cdot \cpo{(X -\rho)} \cd \cpt{(X - \rho)} + \check{f}(X,\rho) \cd \check{U}^{\lambda,k}(X),
\\J^{\lambda}_{2}(X) & =  \min_{ \substack{\rho}} \;  \bset{(X-\rho)  +  [1 - \tpt{X-\rho}] \cd U^{\lambda,1}(X-\rho,\check{\xp}) \nonumber \\ & + \rho \cdot \cpo{(X -\rho)} \cd \cpt{(X - \rho)} + \check{f}(X,\rho) \cd U^{\lambda,2}(X,\check{\xp})} \label{Eq:Jlambda.OneDim.2}
\end{align}
\end{subequations}}
\vspace*{-20pt}
\end{figure*}

For the minimization problem in \eqref{Eq:J.lambda}, using the approximation probabilities in \eqref{Eq:tpo-oneDim} and \eqref{Eq:tptr-oneDim} we first substitute the $\tpz{}$ probabilities with $\ttpz{}$ functions in \eqref{Eq:tJ.lambda}, \eqref{Eq:fi} and \eqref{Eq:gi} and denoting them respectively by  $\cJ^{\lambda}$, $\check{f}_i$ and $\check{g}^\lambda_i$.

With the same approach as in \secref{Ap:DP}, we first start with minimizing $\check{g}^\lambda_i$ as follows:
\begin{align}\label{Eq:Ulambda.0}
U^{\lambda,i}_K(X,\xp) = \min_{ \substack{\rho^{\mc{R}}_{i,k} \in \pi^{\mc{R}}_i\\ \sum_{l=i+1}^K \rho^{\mc{R}}_{i,l} = \xp  \\ \sum_{k=1}^i \rho^{\mc{S}}_{k} = X}} \{ g_i^{\lambda} \}
\end{align}
This will be  in order to find $\check{U}^{\lambda,i}(X) = U^{\lambda,i}_K(X,\check{\xp})$ for different $X$ values, where
\begin{align}\label{Eq:xpHat}
\check{\xp} = \arg_{\xp} \min_{\xp} U_K^{\lambda,i}(X,\xp).
\end{align}

The minimization in \eqref{Eq:Ulambda.0} can be done as follows:
\begin{align}\allowdisplaybreaks \nonumber
U^{\lambda,i}_K(X,\xp)& = \min_{ \substack{\rho^{\mc{R}}_{i,i+1}, \ldots , \rho^{\mc{R}}_{i,K}  \\ \sum_{l=i+1}^K \rho^{\mc{R}}_{i,l} = \xp, \sum_{k=1}^i \rho^{\mc{S}}_{k} = X}} \{ g_i^{\lambda} \} 
 = \min_{ \substack{\rho^{\mc{R}}_{i,K}}} \min_{ \substack{\rho^{\mc{R}}_{i,i+1}, \ldots , \rho^{\mc{R}}_{i,K-1}  \\ \sum_{l=i+1}^{K-1} \rho^{\mc{R}}_{i,l} = \xp - \rho^{\mc{R}}_{i,K}, \sum_{k=1}^i \rho^{\mc{S}}_{k} = X}} \{ g_i^{\lambda} \}
\\\label{Eq:UlambdaK} & = \min_{ \substack{\rho = \rho^{\mc{R}}_{i,K}}} \bset{U^{\lambda,i}_{K-1}(X,\xp -\rho) \;\qquad+ \rho \cdot \cptr(X,\xp -\rho) + \lambda \cdot \cptr(X,\xp)} 
\end{align}
where for $i+3 \leq k \leq K-1$ and for $k = i +2$ we respectively  have \eqref{Eq:Ulambdak} and \eqref{Eq:Ulambda2}.

The minimization process starts with \eqref{Eq:Ulambda2} and then goes on with \eqref{Eq:Ulambdak} and ends with \eqref{Eq:UlambdaK}. The optimization results are stored as $\rho^{\mc{R}}_{i,k}(X,\xp) = \arg_{\rho} U^{\lambda,i}_{k}(X,\xp)$. Therefore, after finding $\check{\xp}$ according to  \eqref{Eq:xpHat}, we find the optimal set of $\check{\rho}^{\mc{R}}_{i,k}(X)$, step-by-step as follows:
\begin{enumerate}
\item $\check{\rho}^{\mc{R}}_{i,K}(X) = \rho^{\mc{R}}_{i,K}(\check{\xp})$
\item for $k: K-1 \rightarrow i+2$
\begin{itemize}
\item $\check{\xp} \leftarrow (\check{\xp} - \check{\rho}^{\mc{R}}_{i,k+1})$
\item $\check{\rho}^{\mc{R}}_{i,k} = \rho^{\mc{R}}_{i,k}(\check{\xp}_k)$
\end{itemize}
\item $\check{\rho}^{\mc{R}}_{i,i+1} = \check{\xp} - \check{\rho}^{\mc{R}}_{i,i+2}$
\end{enumerate}

The next step is to find $\cJ^\lambda$ where  $\cJ^\lambda = \cJ^{\lambda}_K(\check{X_K})$ and, $\check{X} = \arg_{X} \min_{X} J^{\lambda}_K(X)$. The $ \cJ^{\lambda}_K(\check{X_K})$ function is shown as in \eqref{Eq:Jlambda.OneDim} which, in  a recursive form, can be shown  as in \eqref{Eq:Jlambda.OneDim.K}. For $3 \leq k \leq K-1$ we have $\cJ^{\lambda}_k$ as in  \eqref{Eq:Jlambda.OneDim.k}, and for $k = 2$ as shown in  \eqref{Eq:Jlambda.OneDim.2} (According to \eqref{Eq:fi}, $\check{f}(X,\rho)  =\cpt{(X-\rho)} - \cpt{(X)}$).

The minimization process starts with \eqref{Eq:Jlambda.OneDim.2} and then goes on with \eqref{Eq:Jlambda.OneDim.k} and ends with \eqref{Eq:Jlambda.OneDim.K}. The optimization results are stored as $\rho^{\mn{S}}_{k}(X) = \arg_{\rho} J^{\lambda}_{k}(X)$. Then, to find the optimal set of $\check{\rho}^{\mn{S}}_{k}$ we go through the following steps:
\begin{enumerate}
\item $\check{\rho}^{\mn{S}}_{K} = \rho^{\mn{S}}_{K}(\check{X})$
\item for $k: K-1 \rightarrow 2$
\begin{itemize}
\item $\check{X} \leftarrow (\check{X} - \check{\rho}^{\mn{S}}_{k+1})$
\item $\check{\rho}^{\mn{S}}_{k} = \rho^{\mn{S}}_{k}(\check{X})$
\end{itemize}
\item $\check{\rho}^{\mn{S}}_{1} = \check{X} - \check{\rho}^{\mn{S}}_{2}$.
\end{enumerate}

\end{appendices}

\bibliographystyle{IEEEtran}
\balance
\bibliography{IEEEabrv,references_all}{}
\balance



\end{document}